\documentclass[%
 reprint,
 amsmath,amssymb,
 aps,
]{revtex4-2}

\usepackage{graphicx}% Include figure files
\usepackage{dcolumn}% Align table columns on decimal point
\usepackage{bm}% bold math
\usepackage{subfloat}
\usepackage{xcolor}

\newcommand{\be}{\begin{equation}}
\newcommand{\ee}{\end{equation}}
\newcommand{\ba}{\begin{eqnarray}}
\newcommand{\ea}{\end{eqnarray}}
\newcommand{\bd}{\begin{displaymath}}
\newcommand{\ed}{\end{displaymath}}

\def\oneth{{\textstyle{\frac{1}{3}}}}
\def\twoth{{\textstyle{\frac{2}{3}}}}

\begin{document}

\preprint{APS/123-QED}

\title{Comprehensive simulation of heavy-ion collisions at non-zero baryon chemical potential}
% Force line breaks with \\

\author{A. De}
\email{aritrade@umn.edu}
 \affiliation{School of Physics \& Astronomy, University of Minnesota, Minneapolis, MN 55455, USA}%Lines break automatically or can be forced with \\
 
\author{J. I. Kapusta}%
\email{kapusta@umn.edu}
 \affiliation{School of Physics \& Astronomy, University of Minnesota, Minneapolis, MN 55455, USA}%Lines break automatically or can be forced with \\

\author{M. Singh}
\email{singh547@umn.edu}
 \affiliation{School of Physics \& Astronomy, University of Minnesota, Minneapolis, MN 55455, USA}%Lines break automatically or can be forced with \\
 
 \author{T. Welle}
 \email{welle203@umn.edu}
 \affiliation{School of Physics \& Astronomy, University of Minnesota, Minneapolis, MN 55455, USA}%Lines break automatically or can be forced with \\
 
%\date{\taday}% It is always \today, today,

\begin{abstract}
We present results of hydrodynamic modelling of Au-Au collisions from $\sqrt{s_{NN}}$ = 7.7 to 200 GeV. Our simulations have three novel components. Firstly, we use a Linear EXtrapolation of Ultrarelativistic nucleon-nucleon Scattering to nucleus-nucleus collisions (LEXUS) inspired Monte-Carlo initial state model. Secondly, we use a crossover equation of state at finite baryon densities without a critical point. Finally, we use departure functions derived from the quasiparticle theory of transport coefficients for hadronic matter at non-zero baryon densities.
\end{abstract}

\maketitle

%\tableofcontents 

\section{Introduction}

The Beam Energy Scan (BES) at the Relativistic Heavy-Ion Collider (RHIC) made the non-zero baryon chemical potential region of the quantum chromodynamics (QCD) phase diagram accessible to experiments. BES collides gold nuclei at a range of beam-energies and consequently scans different regions of the QCD phase diagram. Experimental programs at the Facility for Antiproton and Ion Research (FAIR) in Darmstadt and at the  Nuclotron-based Ion Collider fAcility (NICA) in Dubna also collide ions at similar energies. These experimental programs complement each other and ensure that a broad region of the QCD phase diagram is covered.

Lattice QCD calculations have shown that the deconfined partons in a cooling quark-gluon plasma (QGP) change phase to a hadron gas by a smooth crossover when the net baryon density is zero \cite{Aoki:2006we,Ding:2015ona,Bazavov:2019lgz}. This is expected for Pb+Pb collisions at the Large Hadron Collider (LHC) and for $\sqrt{s_{NN}}=200$ GeV Au+Au collisions at RHIC. This is in contrast to the first order phase transition expected between the two phases at high baryon densities \cite{Fukushima:2010bq,Fukushima:2013rx,Fischer:2018sdj}. This first-order phase transition line in the QCD phase diagram is expected to end in a critical point where there will be a second-order phase transition. One of the main goals of the BES program is to quantify the location and dynamics of the first and second order phase transitions in QCD. To achieve this goal, we would need phenomenological modelling of these experiments to extract relevant physics insights from these experiments.

Significant progress has been made towards phenomenological description of high-energy heavy ion-collisions where net baryon densities are close to zero. References \cite{Heinz:2013th,Braun-Munzinger:2015hba,Romatschke:2017ejr,Shen:2020mgh} provide recent reviews on modelling ultrarelativistic heavy-ion collisions.  The standard model of heavy-ion collisions is comprised of a pre-hydrodynamic phase, a hydrodynamic phase, and a post hydrodynamic phase. The pre-hydrodynamic phase is usually modeled by approximating the colliding nuclei as being infinitely thin in the beam direction \cite{Romatschke:2009im,Ollitrault:2007du}. The is justified because of significant Lorentz contraction. A variety of initial state models are in popular use for these high energy collisions \cite{Miller:2007ri,Drescher:2006ca,Drescher:2007ax,Krasnitz:1999wc,Krasnitz:2000gz,Pang:2012he,Schenke:2012wb,Moreland:2014oya}. The energy density profile from the pre-hydrodynamic phase provides the initial conditions for the hydrodynamic phase. The expanding matter is simulated by relativistic viscous fluid dynamics. Apart from the energy and momentum conservation equations of the hydrodynamics, one also needs to know the equation of state (EOS) of the nuclear matter. Typically, in the high energy regime, the EOS is obtained by matching the high temperature EOS from lattice QCD with low temperature EOS from hadron resonance gas models using a parameterization \cite{Huovinen:2009yb,Moreland:2015dvc} or a switching functions \cite{Albright:2014gva}. As the fluid expands, the mean free paths become large and hydrodynamics is no longer an appropriate theory for describing the system. At this stage we switch to the kinetic theory description. Hadron distributions are obtained from the fluid by matching the energy-momentum tensor of the fluid with the energy-momentum distribution function of hadrons using the Cooper-Frye prescription \cite{Cooper:1974mv}. Viscous corrections to the fluid energy-momentum tensor are matched to the non-equilibrium corrections to the particle distribution functions using departure functions. The hadrons produced collide with each other and resonances decay, eventually leading to the chemical freezeout followed by the kinetic freezeout, at which point all the particles free stream to detectors.

Extending the standard model of heavy-ion collisions to BES energies with non-zero baryon densities poses challenges. A recent report by the BEST collaboration discusses these issues in detail \cite{An:2021wof}. We will deal with some of the issues here.

The assumption of ultrarelativistic colliding nuclei being infinitely thin no longer holds as the collision energies are much lower. Consequently, the initial state dynamics changes from being approximately two-dimensional to being fully three-dimensional. This necessitates dynamical initialization \cite{Shen:2017bsr,Du:2018mpf,Shen:2017ruz,Akamatsu:2018olk} of hydrodynamics. For this, we need a 3+1D space-time and momentum distribution of initial energy and charge sources. The assumption of longitudinal boost-invariance is no longer a good approximation. There are some initial state models which deal with these issues \cite{Shen:2020jwv,Shen:2017bsr,Okai:2017ofp,Shen:2017ruz,Du:2018mpf}.

Here we propose a new LEXUS-inspired 3D initial state. LEXUS stands for Linear EXtrapolation of Ultrarelativistic nucleon-nucleon Scattering to nucleus-nucleus collisions; the model was introduced in Ref. \cite{Jeon:1997bp}. LEXUS treats nucleus-nucleus collisions as superposition of individual nucleon-nucleon collisions. It uses parameterized data from nucleon-nucleon collisions and extrapolates it to describe nucleus-nucleus collisions. LEXUS was originally formulated in momentum space. Here we use a Monte-Carlo sampling of nucleon positions and treat nuclear collisions as a sequence of binary nucleon-nucleon collisions to obtain spatial information. Energy loss in each binary collision is given by a distribution from LEXUS which is fit to nucleon-nucleon collision data. The idea is to use known results from nucleon-nucleon collisions to fix the free parameters of the model.

As the baryon charge densities are non-zero, one needs to keep track of an additional $U(1)$ charge conservation along with the usual energy-momentum conservation of fluid dynamics. Also, an equation of state is required where the thermal quantities are also a function of baryon potential in addition to being functions of temperature. Some progress has been made in this direction recently. These calculations usually extrapolate lattice calculations to Taylor expansion coefficients in some order of chemical potential over temperature \cite{Hegde:2014sta,Guenther:2017hnx,Bazavov:2017dus,Monnai:2019hkn,Noronha-Hostler:2019ayj,Borsanyi:2021sxv}. In this work we employ a crossover EOS \cite{Albright:2014gva} which matches parameterized perturbative QCD EOS at high temperatures without a critical point to hadron resonance gas EOS at low temperatures using a switching function.

Departure functions need modifications to account for the additional $U(1)$ charge. They have largely been calculated at zero baryon densities \cite{Teaney:2003kp,Bozek:2009dw}. We calculate and implement departure functions calculated using quasiparticle theory \cite{Albright:2015fpa} at finite baryon chemical potentials and within the relaxation time approximation.

This is a comprehensive heavy ion collision simulation at finite baryon chemical potential meant to be seen as a baseline study. The organization of this paper is as follows. The initial state model is given in Sec \ref{Sec:Initial_State}. We begin with a short summary of the original LEXUS paper. This helps in understanding the various model choices we make in the LEXUS inspired 3D initial state. We discuss the EOS used in Sec. \ref{Sec:EOS}. We calculate the departure functions at finite baryon chemical potentials in Sec. \ref{Sec:Departure_Functions}. Finally, we give our comparisons with data in Sec. \ref{Sec:Results} and our conclusions in Sec. \ref{Sec:Conclusions}.

\section{LEXUS inspired 3D initial state}\label{Sec:Initial_State}

\subsection{Summary of the original LEXUS model}
We begin by giving a brief summary of the original LEXUS model \cite{Jeon:1997bp}. Knowledge of the original LEXUS paper is not required to understand this paper but it helps to understand the various motivations and inspirations of the LEXUS based 3D model proposed in this paper.

In the absence of an ab-initio QCD calculation for nuclear collisions, LEXUS served as a baseline study compared to other phenomenological models of initial states. When LEXUS was first published, it described the rapidity and transverse momentum distributions of baryons in central sulfur-sulfur and lead-lead collisions at the SPS pretty well. LEXUS relies only on the data from nucleon-nucleon collisions which is extrapolated to form a model of nucleus-nucleus collisions.

The salient features of LEXUS are as follows

\begin{itemize}
    \item In LEXUS, nucleons are arranged in rows. When the projectile and the target rows pass through each other, all the nucleons in one row collide with all the nucleons in the other row. The main object of interest in LEXUS is the two particle rapidity distribution $W_{mn}$ for the $m$th projectile nucleon and $n$th target nucleon immediately after their collision. The $W_{mn}$ is a result of a collision of a projectile nucleon, which has already undergone $n-1$ previous collisions and a target nucleon, which has undergone $m-1$ previous collisions. For a projectile nucleon with rapidity $y'_P$ colliding with a target nucleon of rapidity $y'_T$ resulting in two nucleons with rapidities $y_P$ and $y_T$, we have
    \begin{eqnarray}
    \hspace{20pt} W_{mn} (y_P, y_T) &=& \int dy'_P dy'_T W^P_{mn-1}(y'_P) W^T_{m-1n}(y'_T) \nonumber \\
    &\times& K(y'_P+y'_T \rightarrow y_P+y_T).
    \end{eqnarray}
    Here $y$ is the momentum rapidity. The collision kernel $K$ is chosen to be Markovian. In the original LEXUS model, this kernel was chosen to be
    \begin{eqnarray}
        K(y'_P+y'_T \rightarrow y_P+y_T) &=& \lambda K_{\text{inelastic}} \nonumber \\
        & +& (1-\lambda)K_{\text{elastic}},
    \end{eqnarray}
    where
    \begin{eqnarray}\label{eq:lexus_kernel}
    K_{\text{inelastic}} = \frac{\cosh(y_P-y'_T)}{\sinh(y'_P-y'_T)}\frac{\cosh(y'_P-y_T)}{\sinh(y'_P-y'_T)} 
    \end{eqnarray}
    and
    \begin{eqnarray}
     K_{\text{elastic}} = \delta(y'_P-y_P)\delta(y'_T-y_T).
    \end{eqnarray}
    The $\cosh$ function is chosen for the distribution because, in a high-energy nucleon-nucleon collision, the distribution of outgoing nucleons is flat in longitudinal momentum (or, in other words, hyperbolic cosine in rapidity). The coefficient $\lambda$ is the fraction of nucleon-nucleon collisions that are inelastic and non-diffractive.
    \item LEXUS arrives at the following expression for the final baryon rapidity distribution arising from the projectile participants
    \begin{eqnarray}
    \frac{dN_P}{dy}(y,\mathbf{b}) &=& \sum_{\bar{m}=1}^{A_P} \sum^{\bar{m}}_{m=1} \sum_{n=1}^{A_T} W^P_{mn}(y) \nonumber \\ 
    &&\times  \int \frac{d^2 s_P}{\sigma_{NN}} \mathcal{P}_n^T(\mathbf{s}_T)\mathcal{P}_{\bar{m}}^P(\mathbf{s}_P)
    \end{eqnarray}
    Here $\sigma_{NN}$ is the nucleon-nucleon cross-section which was chosen to be a constant, and the $s_P$ and $s_T$ refer to the transverse positions.  There is a symmetrical contribution from the target participants. The $\mathcal{P}$ was chosen to be a binomial distribution, subject to the condition that the nucleons follow the Woods-Saxon distribution.
    \item Transverse momentum is put into LEXUS in the form of a random walk in the transverse momentum space. The average transverse momentum squared of a baryon after $k$ collisions is $k\langle p_T^2\rangle_{NN}$, where $\langle p_T^2\rangle_{NN}$ is the average in a nucleon-nucleon collision.
    \item There was no hydrodynamic evolution used in the original LEXUS model. The rapidity and transverse momentum distributions of baryons obtained from this model compared favorably to experimental data available at the time. There were comparisons made for multiplicity of negatively charged hadrons and their rapidity and transverse momentum distributions.
\end{itemize}

\subsection{LEXUS-inspired 3D initial state}
 As in LEXUS, we assume that the nucleons follow straight-line trajectories, striking nucleons from the other nucleus that lie in their path and interacting with them like two nucleons will interact in free space. The model does not distinguish between neutrons and protons. The original LEXUS model was formulated in momentum space but for our Monte Carlo model, we need to specify the coordinate space information. 

We sample the nucleonic positions randomly from a Woods Saxon distribution. 
\begin{equation}
    \rho (r) = \frac{\rho_0}{1+\mathrm{exp}[(r-R)/a]}
\end{equation}
Here $R$ is the nuclear radius, $\rho_0$ is the density at the center of the nucleus, and $a$ is the nuclear skin-thickness. Once the nucleonic positions are specified, Lorentz contraction is applied in the beam direction. We initialize all the nucleons with the beam rapidity and no transverse motion. In the future, this can be easily generalized to include transverse flow. If the two nuclei are moving with opposite velocities of $v_z$ and the radius of the nucleus is $R$, the overlap time of the two nuclei is
\begin{equation}
    \tau_{\text{overlap}} = \frac{2 R}{\gamma v_z} = \frac{2R}{\sinh(y_{\text{beam}})}.
\end{equation}
Here $\gamma$ is the Lorentz factor and $y_{\text{beam}} = \text{arccosh}(\sqrt{s_{NN}}/(2m_N))$ is the beam rapidity. The collision energy per nucleon pair is $\sqrt{s_{NN}}$. The mass of the proton is $m_N = 0.938$ GeV. Longitudinal thickness in the overlap leads to a considerable overlapping time of $\tau_{\text{overlap}}\sim 2-3$ fm for lower BES energies. The target and the projectile nucleons are initialized with velocities
\begin{eqnarray}
    v_z^{\text{Projectile}} = \tanh(y_{\text{beam}}), \\
    v_z^{\text{Target}} = -\tanh(y_{\text{beam}}).
\end{eqnarray}
For the target moving to the left and the projectile moving to the right, we set $\mathrm{max}\{z_{i}^{\text{Projectile}}\} = \mathrm{min}\{z_{j}^{\text{Target}}\} = 0$ at time $t = 0$. In other words, we set the zero of time to be $t = 0$ at the moment when the right-most projectile nucleon crosses the left-most target nucleon. The longitudinal position of this nucleon pair is defined to be $z = 0$. This particular nucleon pair may or may not collide depending on their positions in the transverse reaction plane. The nucleons travel in a straight line and collide with other nucleons within a fixed non-diffractive inelastic scattering cross section. We use the geometric interpretation of the nucleon-nucleon cross section and the transverse positions of the nucleons to determine whether a collision takes place. The total nucleonic scattering cross section is chosen to be a constant 42 mb, which is approximately the average of the total nucleon cross section for the collision energies considered in this paper. We only accept $60 \%$ of the collisions because that is the percentage of total collisions that are inelastic and non-diffractive \cite{Videbaek:1995mf}. Only the non-diffractive inelastic collisions source into hydrodynamic evolution.

 The model assumes that the nucleons from one nucleus strike nucleons from other nucleus that lie in their path and interact with them exactly as they would in free space. The longitudinal coordinate and time of binary collision is determined by the space-time location of the nucleons crossing each other with $z=0$ and $t=0$ defined above. The transverse coordinates are defined as
 \begin{eqnarray}
 x_{\text{binary-collision}} = (x_P + x_T)/2, \\
 y_{\text{binary-collision}} = (y_T + y_T)/2.
 \end{eqnarray}
 Here $(x_T, y_T)$ and $(x_P, y_P)$ are the transverse positions of the participating target and projectile nucleons, respectively.
 
\begin{figure}[]
\centering
\includegraphics[scale=0.08]{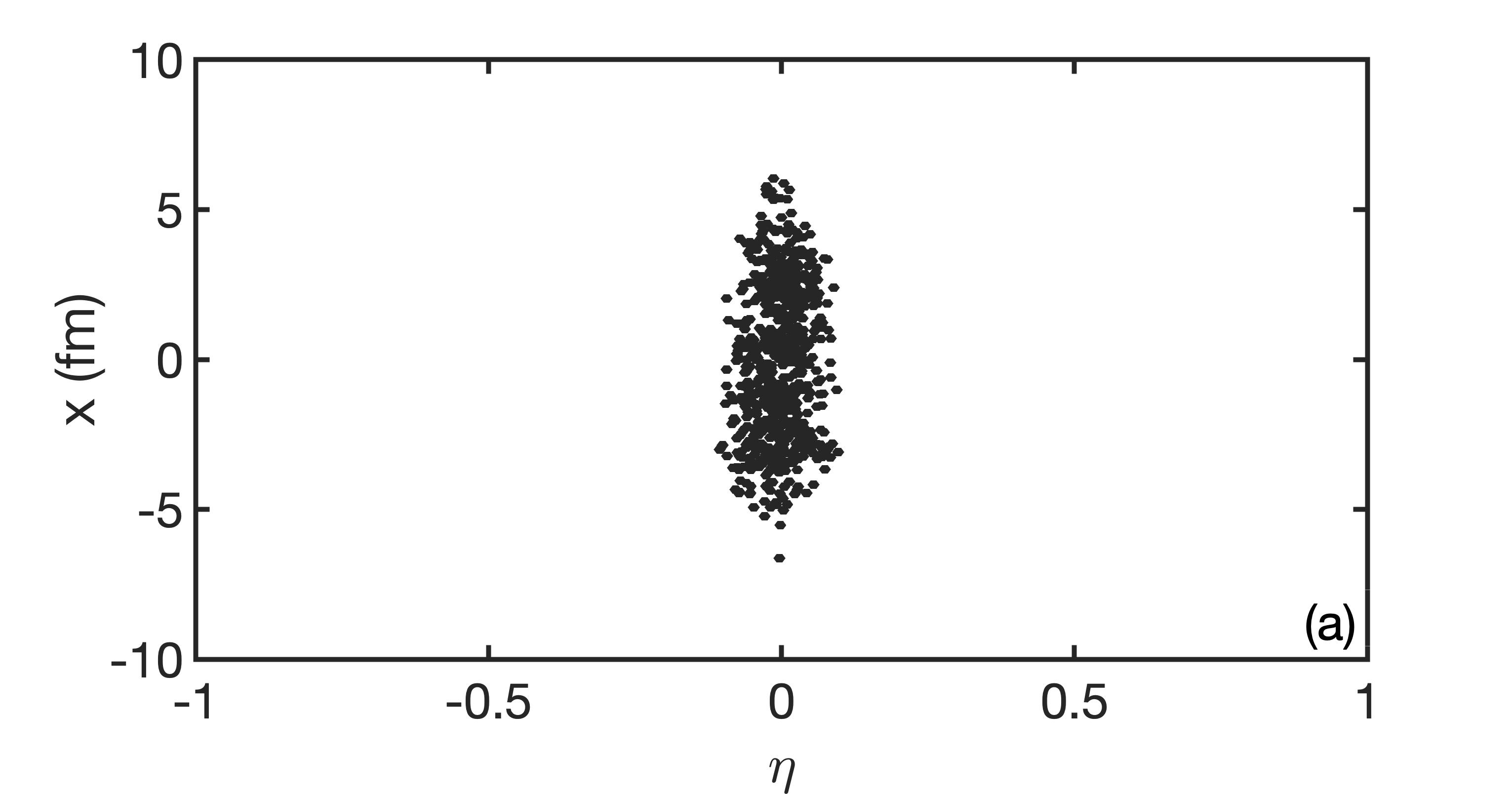}
\includegraphics[scale=0.08]{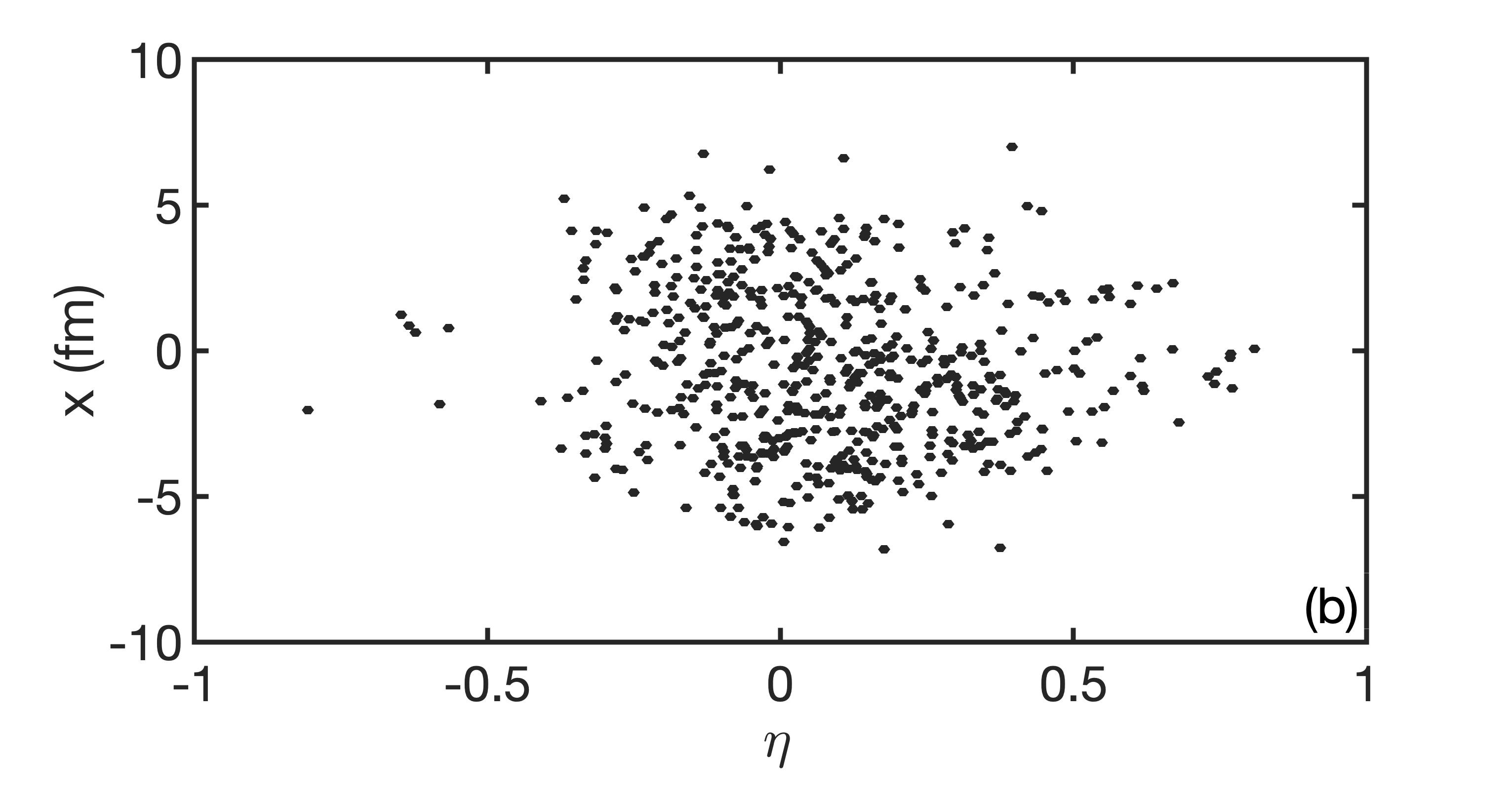}
\caption{Binary collision locations for a Au + Au collision at $\sqrt{s_{NN}} = 200$ GeV (a) and at $\sqrt{s_{NN}} = 11.5$ GeV (b).}
\label{fig:binary_collisions}
\end{figure}

Participating nucleons continue on their trajectory with reduced momentum from the point of binary collision. Nucleons can undergo multiple collisions and can even reverse direction if it they have undergone significant momentum change from a collision. These binary collision positions can be turned into space-time rapidity coordinates. Figure \ref{fig:binary_collisions} shows the binary collision positions for a $\sqrt{s_{NN}} = 200$ GeV and a $\sqrt{s_{NN}} = 11.5$ GeV Au+Au collision, at zero impact parameter. One can see the effect of Lorentz transformation in the spread in space-time rapidity $\eta = \tanh^{-1}(z/t)$ of the binary collision locations.  There is greater Lorentz contraction at $\sqrt{s_{NN}} = 200$ GeV, consequently binary collision locations occupy a narrower region in space-time rapidity $\eta$.
  
Let us now describe the rapidity and proper time dependence of the energy and net baryon density deposition. The energy lost in a single binary collision is sampled from the probability distribution
\begin{equation}
P(y_{\text{loss}}) = \frac{\cosh(2y^{\text{total}}_{\text{rest-frame}}-y_{\text{loss}})}{\sinh(2y^{\text{total}}_{\text{rest-frame}}) - \sinh(y^{\text{total}}_{\text{rest-frame}})}.
\label{eq:yloss_distribution}
\end{equation}
 The absolute value of the incoming nucleon's rapidity in the pair rest frame is $y^{\text{total}}_{\text{rest-frame}}$ and the rapidity loss in a given binary collision is $y_{\text{loss}}$. The probability distribution is chosen to be a $\cosh$ distribution as explained in the previous section and is inspired from Eq. (\ref{eq:lexus_kernel}). The relation between Eq. (\ref{eq:lexus_kernel}) and the above distribution is explained in appendix \ref{App:lexus_comparison}. The distribution is normalized to 1. This distribution function is approximately the same distribution function that was used in the original LEXUS paper. Such distributions have also been used in Refs. \cite{Shen:2017bsr,Hwa:1983ik,Csernai:1984vs}. We sample the rapidity lost $y_{\text{loss}}$ in the range $[ 0,y^{\text{total}}_{\text{rest-frame}}]$, where \begin{equation}
      y^{\text{total}}_{\text{rest-frame}} = \left|y_{\text{target}}\right| + \left|y_{\text{projectile}}\right|.
  \end{equation}
  
  \begin{figure*}[]
  \centering
  \begin{tabular}{cc}
   \includegraphics[scale=0.075]{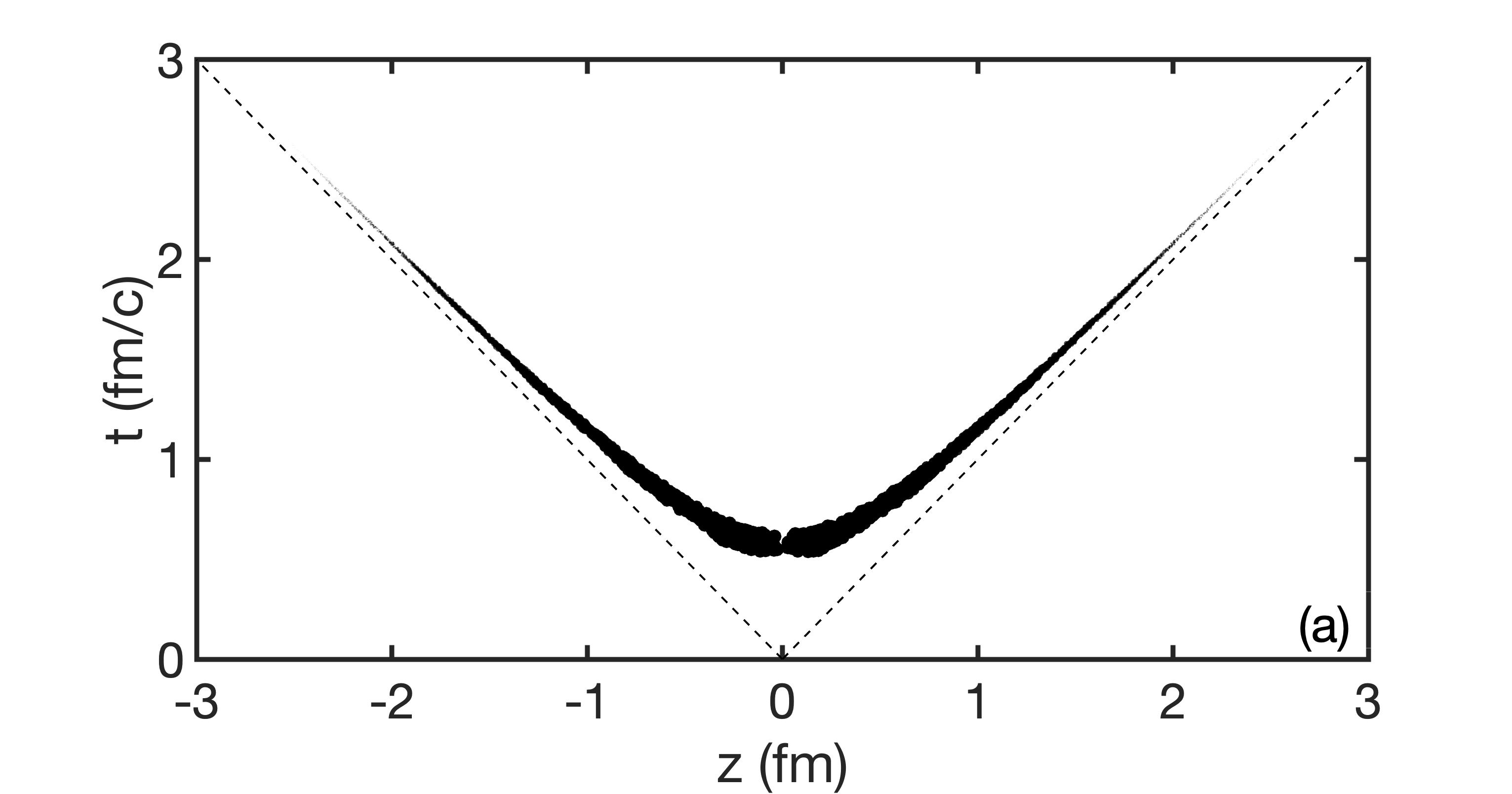} & 
   \includegraphics[scale=0.075]{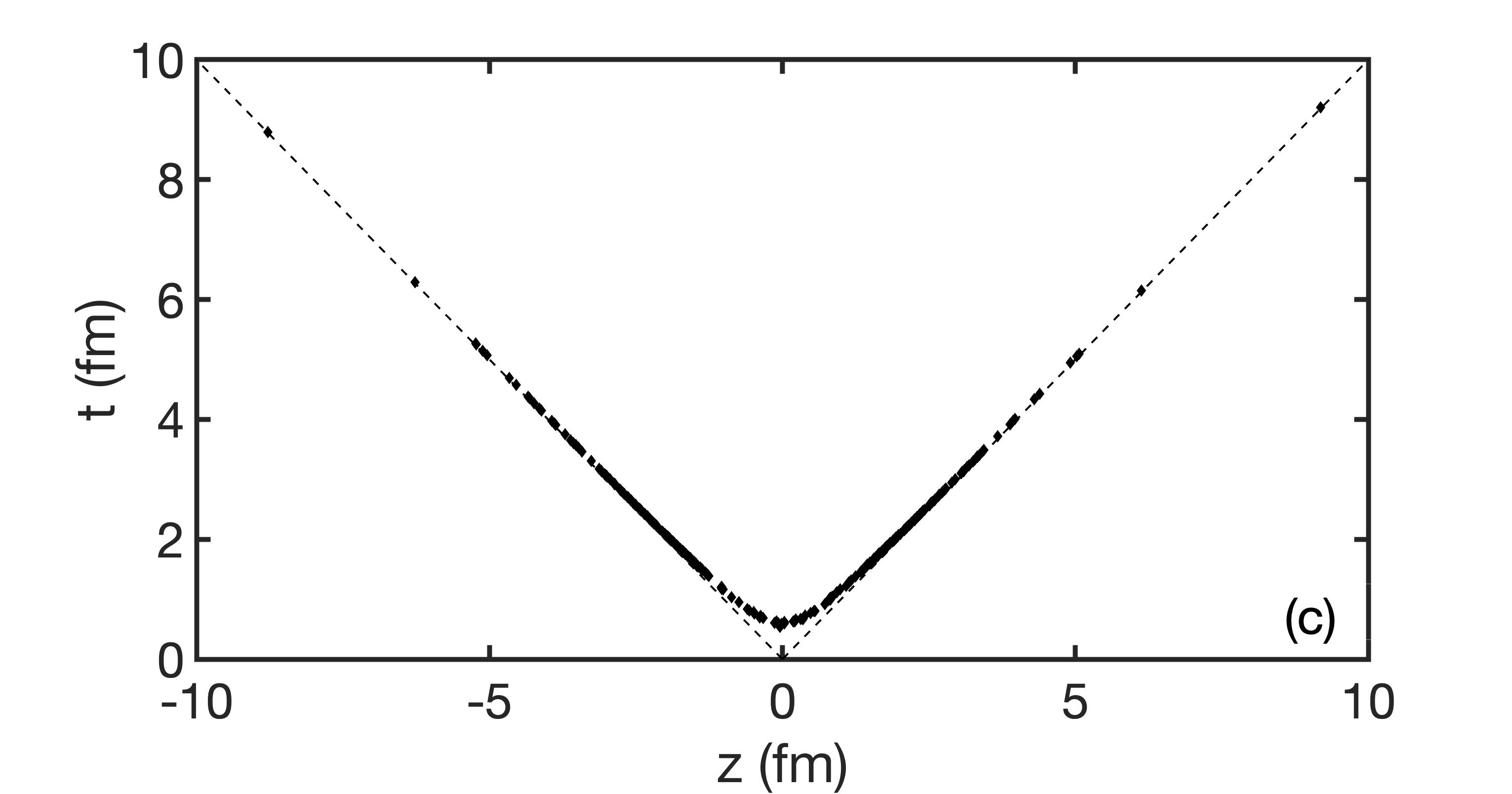} \\
  \includegraphics[scale=0.075]{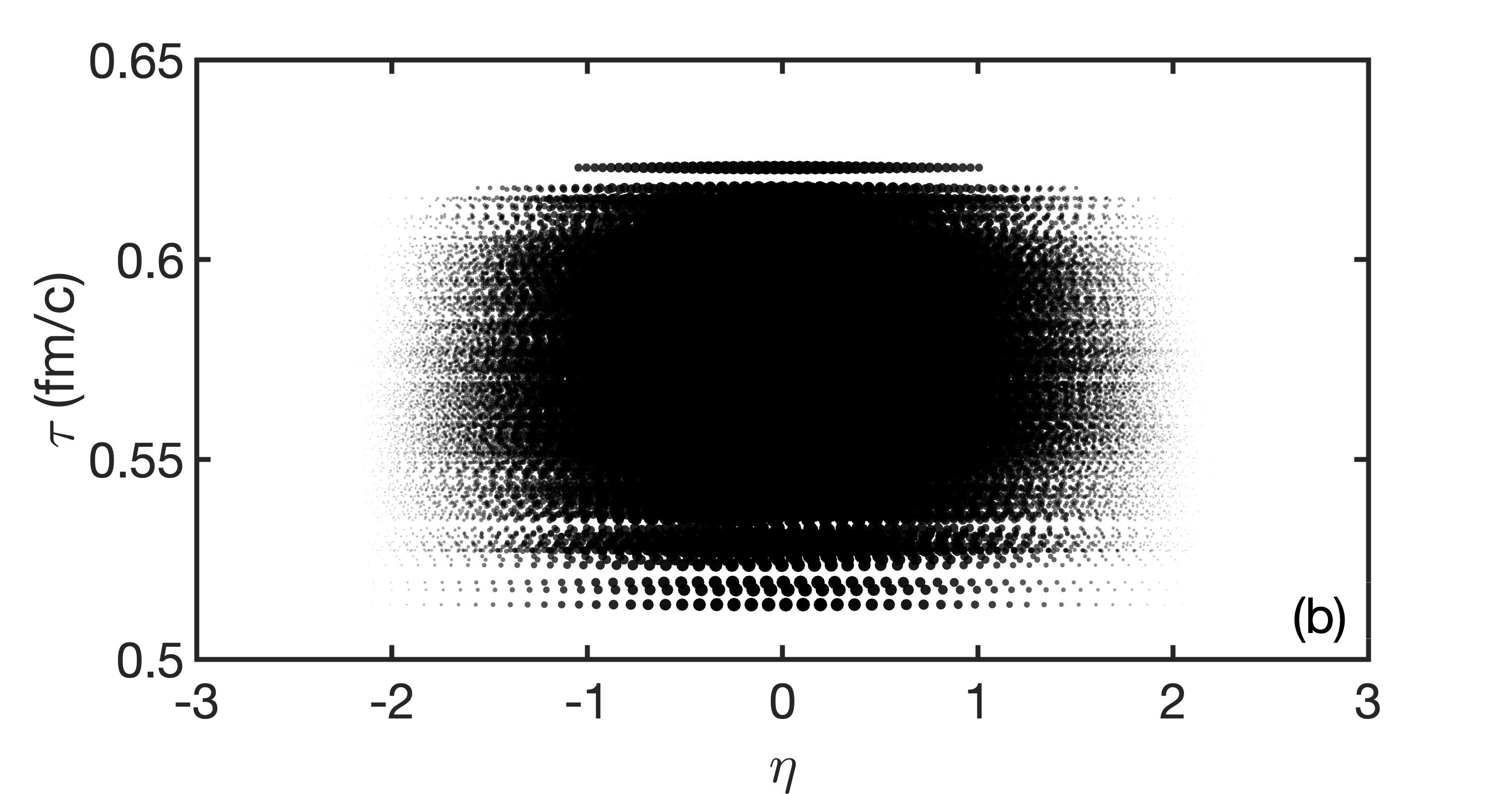} &
   \includegraphics[scale=0.075]{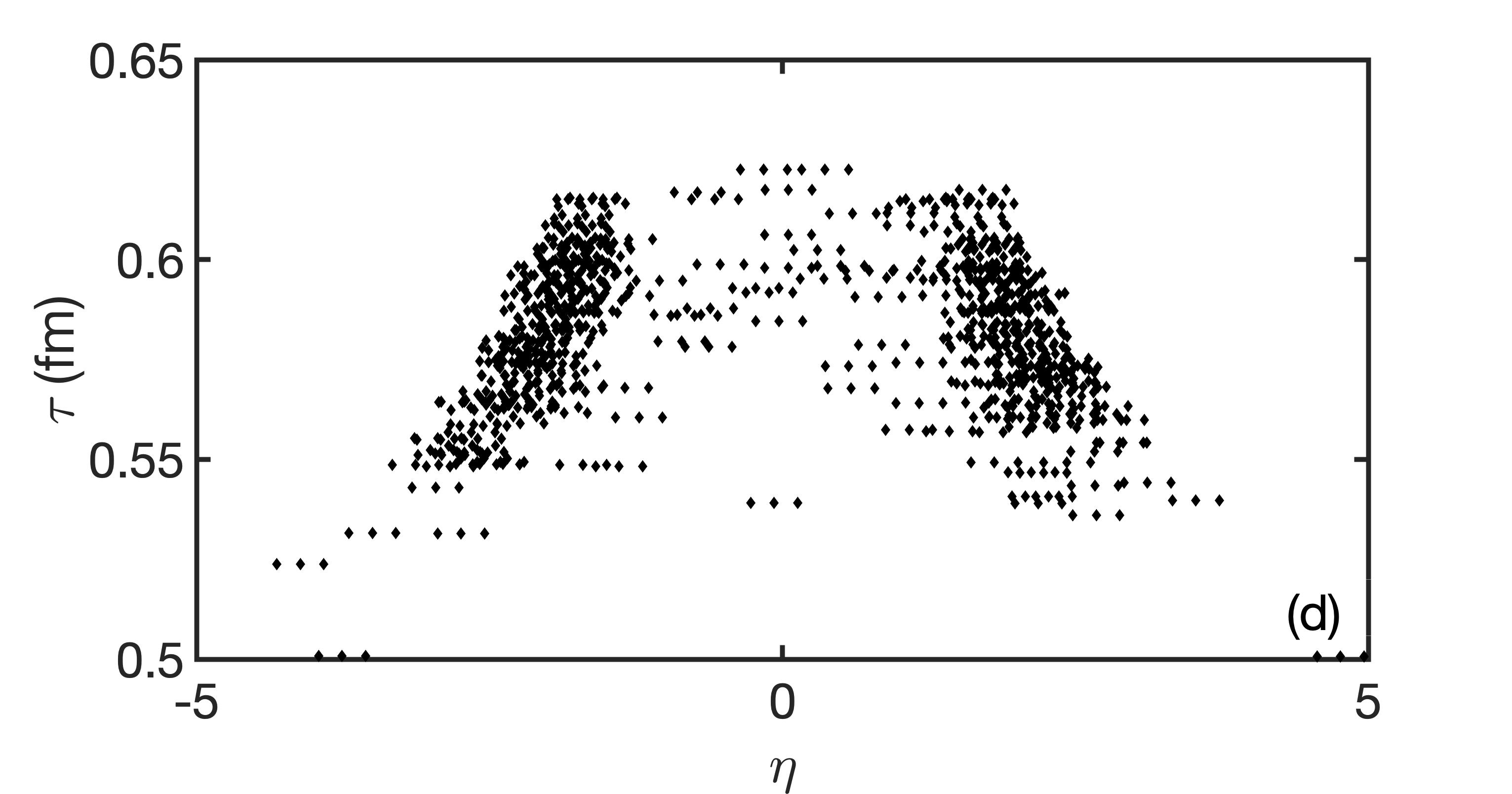}
   \end{tabular}
  \caption{Binary collision energy deposition in the Cartesian (a) and Milne (b) coordinates. Participant nucleon locations in the Cartesian (c) and Milne (d) coordinates. The plots are for a zero impact parameter $\sqrt{s_{NN}} = 200$ GeV Au+Au collision.}
  \label{fig:200GeV_gaussians_remnants}
\end{figure*}

\begin{figure*}[]
  \centering
  \begin{tabular}{cc}
   \includegraphics[scale=0.075]{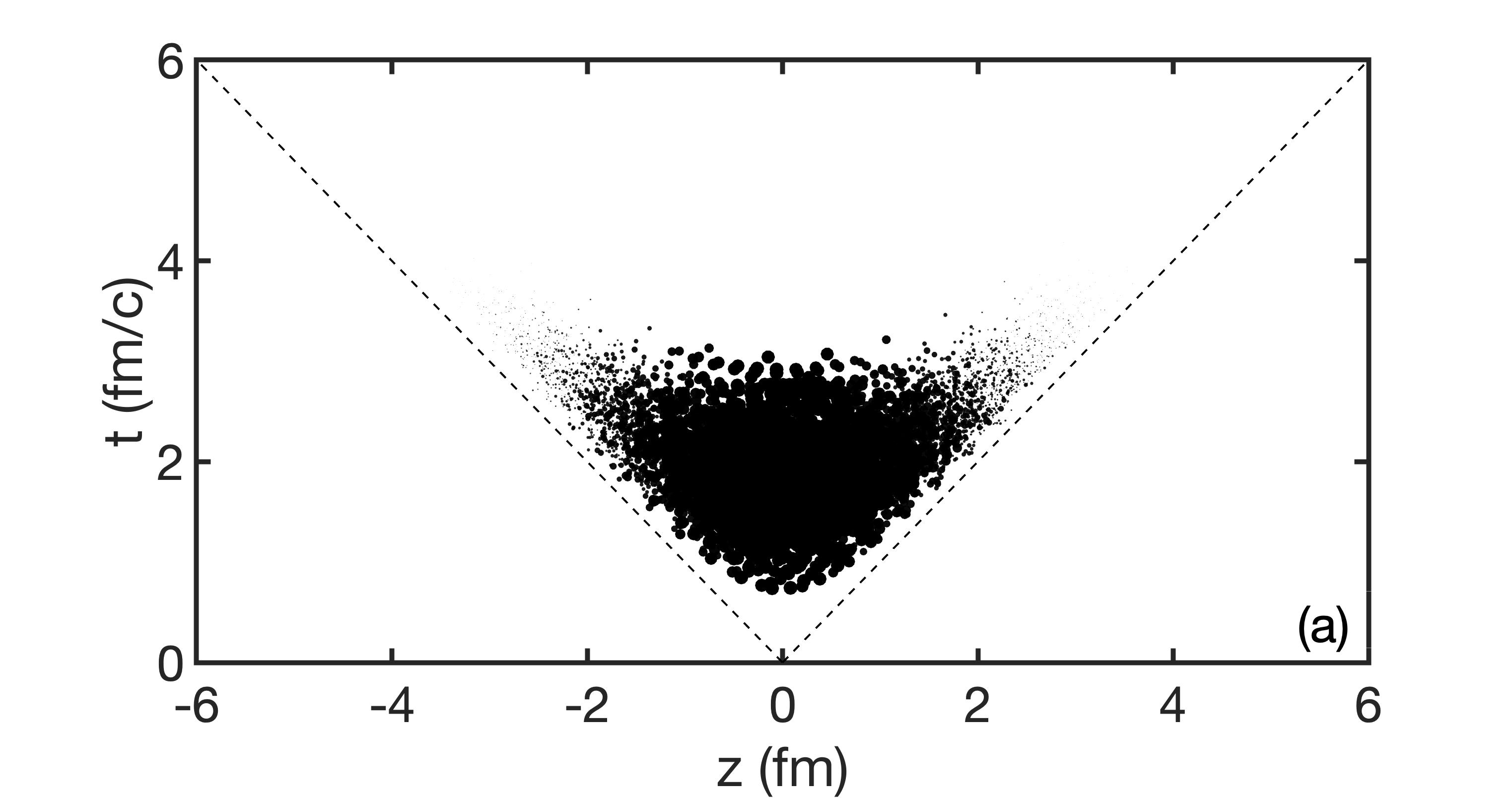} & 
   \includegraphics[scale=0.075]{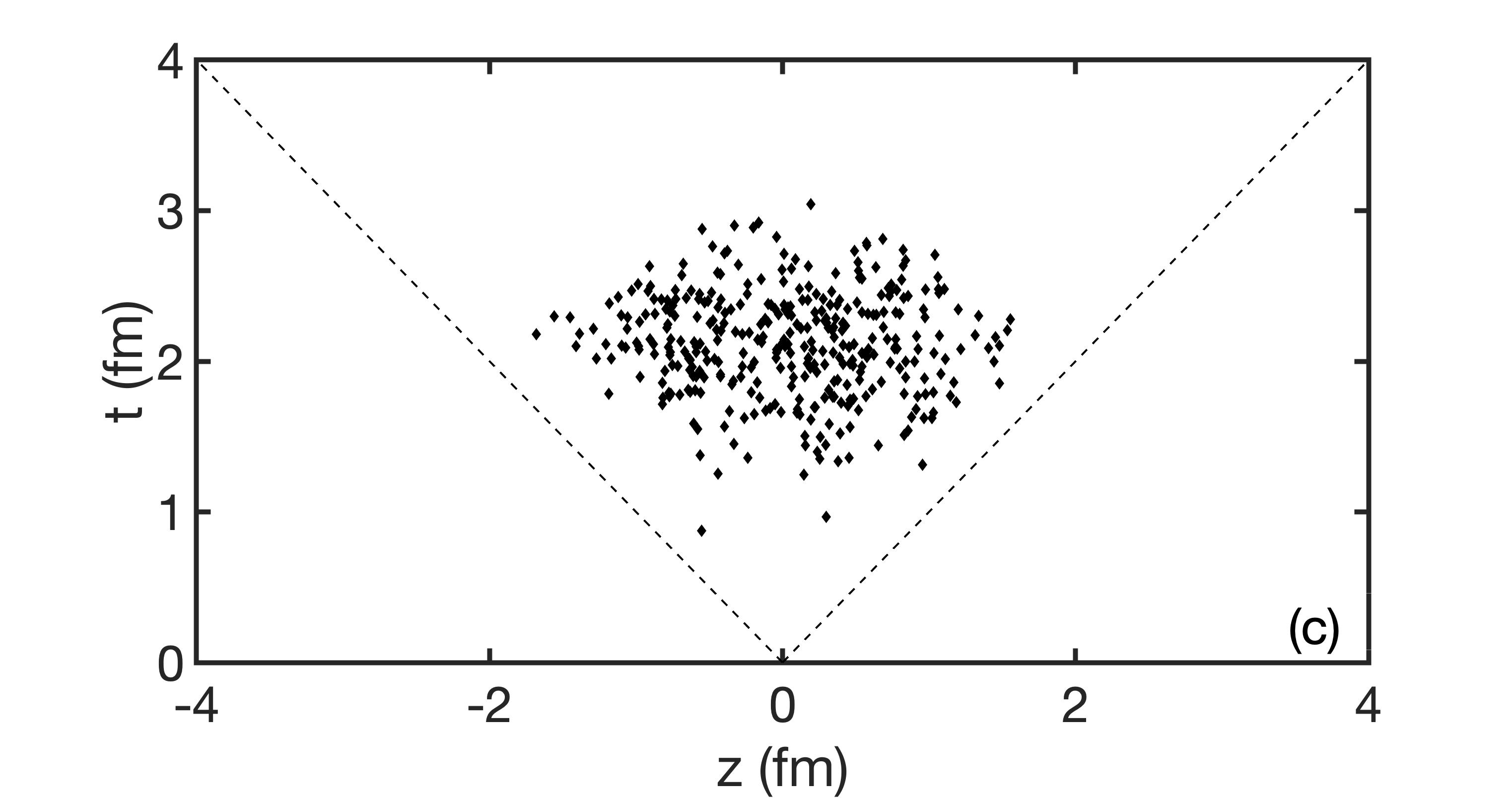} \\
  \includegraphics[scale=0.075]{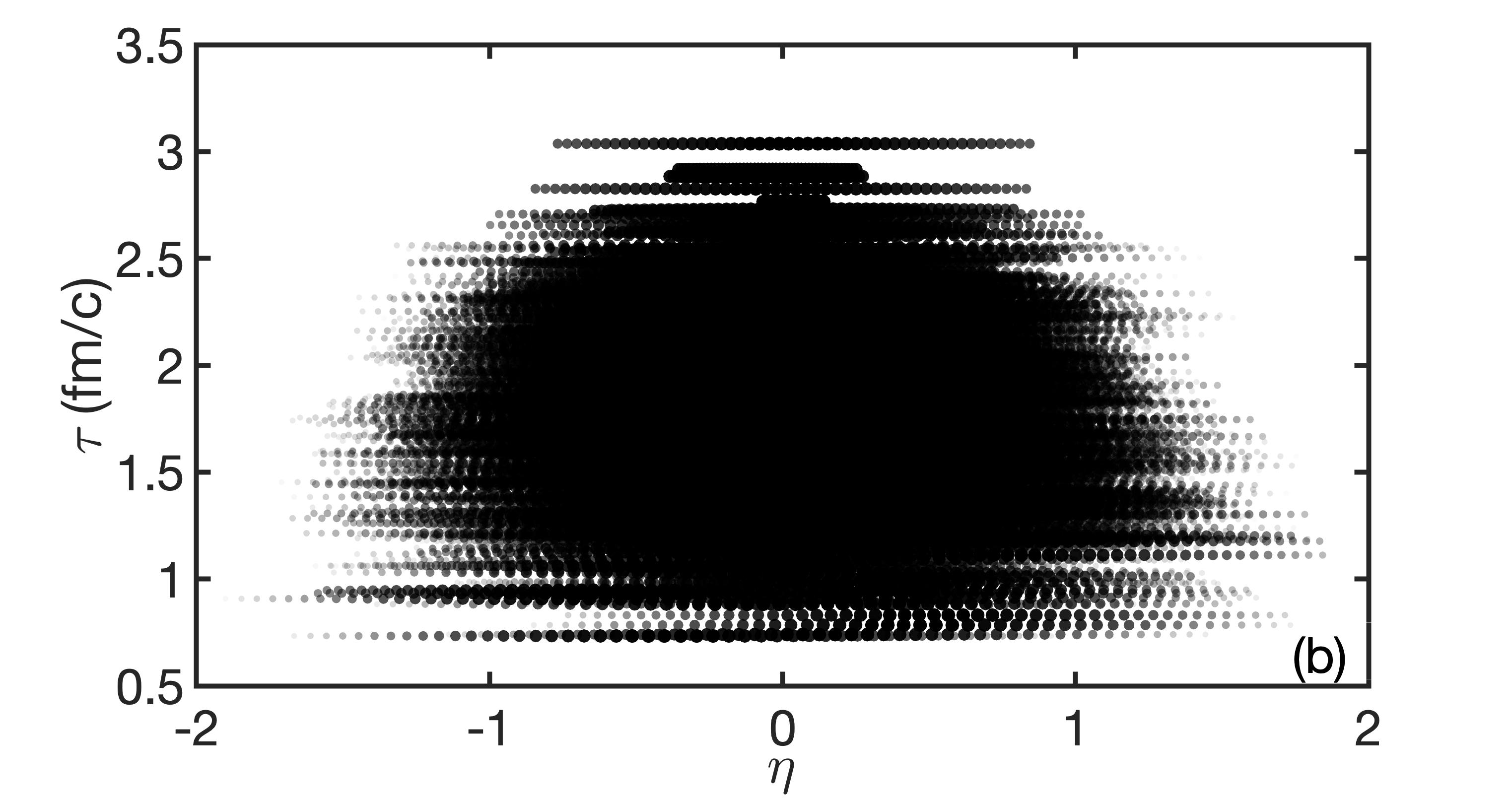} &
   \includegraphics[scale=0.075]{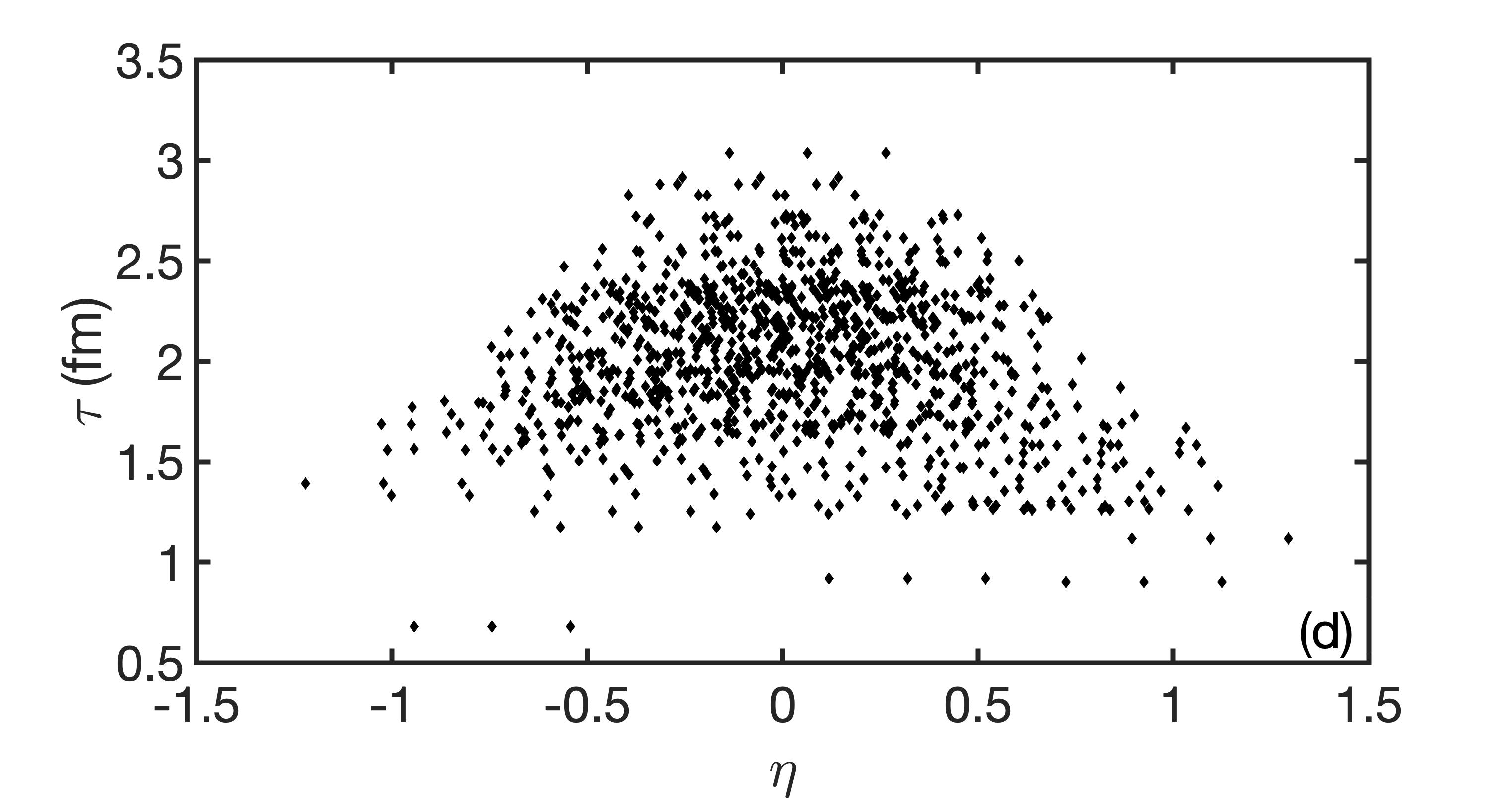}
   \end{tabular}
  \caption{Same as fig. \ref{fig:200GeV_gaussians_remnants} but for a $\sqrt{s_{NN}} = 11.5$ GeV Au+Au collision.}
  \label{fig:11pt5GeV_gaussians_remnants}
\end{figure*}

The rapidity lost in a single binary collision $y_{\text{loss}}$ determines the amount of energy sourced into the hydrodynamic equations. The energy that is deposited in the binary collision is distributed as a Gaussian in rapidity space. We make this choice because the rapidity distribution of particles produced in nucleon-nucleon collisions is approximately Gaussian as evidenced in the Landau model \cite{Florkowski:2010zz}. The width is 
\begin{equation}
    \sigma_{\eta}^2 = \frac{8}{3} \frac{c_0^2}{1-c_0^4} \ln \left( \frac{\sqrt{s_{b}}}{2m_N} \right)=\ln \left(\frac{\sqrt{s_{b}}}{2m_N} \right)
    \label{gaussian_sigma}
\end{equation}
Here $\sqrt{s_{b}}$ signifies the center of mass energy in a particular binary collision. The speed of sound squared $c_0^{2} = 1/3$ is a good fit for nucleon-nucleon collision data \cite{Carruthers:1973ws}. The energy deposition from the collisions in the hydrodynamic phase occurs after a fixed time $t_{\text{thermalize}} = 0.5$ fm. The energy deposition at $\{ \tau_c,\eta_c,x_c,y_c\}$ with a width of $\sigma_{\eta}$ is instantaneous in the $\tau, \eta$ coordinate. The original LEXUS model has no free parameters. Here, in the LEXUS inspired 3-D initial state, the thermalization time for deposited energies and participants is the only free parameter. Importantly, there is no normalization factor associated with the initial distribution of energy. Once the two-particle $y_{\text{loss}}$ is sampled, it is equally subtracted from the rapidity of the colliding nucleons before the collision. That means that even if the target nucleus is moving to the left and projectile nucleus is moving to the right initially, depending on the magnitude of $y_{\text{loss}}$ sampled in any given collision, the projectile might up end up moving to the left or the target to the right.

Once a nucleon has undergone a collision and has deposited some of its energy, it is called a participant. Participants can undergo further collisions. We assume that all participants, after they have undergone their last binary collision, will deposit their energy and baryon charge into the quark-gluon plasma. The nucleons that have not taken part in any binary collisions, called spectators, are allowed to pass through. The energy and baryon charge from participants are deposited after they are propagated for $\Delta \tau = 0.5$ fm after their last collision. This means that the participants will have propagated to different positions (based on their individual energies) before being absorbed into the produced matter.

Figures \ref{fig:200GeV_gaussians_remnants} and \ref{fig:11pt5GeV_gaussians_remnants} show the distribution of collision energy deposition and participant locations in laboratory $t-z$ and Milne $\tau-\eta$ coordinates for Au+Au collisions at 200 GeV and 11.5 GeV, respectively, at zero impact parameter. The left-hand plots in both Figures \ref{fig:200GeV_gaussians_remnants} and \ref{fig:11pt5GeV_gaussians_remnants} represent the positions of the binary collision energy distributions for the two collision energies. The right-hand plots of the same figures represent the positions of participants for the two beam energies. One can see the effects of Lorentz contraction in the spread in the space-time rapidity and $z$ coordinate of the binary collision energy depositions and participants. For the 200 GeV collision, there is a wider interval in rapidity around midrapidity, where binary collision energy deposition is independent of rapidity. This is a check that at higher energies, the Bjorken model is a more accurate approximation. This is also visible when one compares the $t-z$ plots for the two energies. The extent of the deposition regions depicted in the plots is $1 \sigma$ of the aforementioned Gaussian distribution mentioned in Eq.~\eqref{gaussian_sigma}. One can see the fading of the plots at the edges of these deposition regions, which just signifies the tailing of energy depositions. The $\tau-\eta$ plots start from $\tau = 0.5\, \mathrm{fm}/c$, which is a consequence of the parameter choice of $\tau_{\text{thermalize}}$ we explained earlier.

The different participant position profiles for the two energies is due to the difference in energies. For 200 GeV collisions, even after binary collisions, nucleons possess considerable energy that could be one or two orders of magnitude greater than their rest mass. Consequently, they still traverse at speeds close to the speed of light and graze the light cone as is evident in the figure. On the other hand, for 11.5 GeV collisions, participant nucleons possess energy of the order of their rest mass and therefore travel at much lower velocities. This is why they are more uniformly spread in the rapidity direction.

\subsection{Initializing Hydrodynamics}

We use the publicly available hydrodynamic solver MUSIC \cite{Schenke:2010nt}. In this section, we describe how we initialize our hydrodynamical equations.

Each binary collision energy deposition carries the following information in order to fully quantify the source terms for the hydrodynamic evolution.
\begin{equation}
\tau_c, x_c, y_c, \eta_c, \sigma_{\eta}, E, p_z    
\end{equation}
Here $E$ is the energy deposited and $p_z$ is the momentum of the center of mass of the binary collision.
Participants have the following information.

\begin{equation}
    \tau, x, y, \eta, E, p_z
\end{equation}
Here $E$ is the remaining energy of the nucleon and $p_z$ is the momentum with which it was traveling at the time it was dissolved into the dense matter.  Each of them has a baryonic charge of $+1$.

These baryon energy and participant depositions act as sources for the hydrodynamic equations
\begin{eqnarray}\label{eq:hydro_em}
    \partial_{\nu}T^{\mu\nu} &=& S^{\mu}_{\mathrm{source}}(\tau,\mathbf{x}),\\\label{eq:hydro_b}
    \partial_{\mu}J^{\mu}_{B} &=& \rho_{B,\mathrm{source}}(\tau,\mathbf{x}).
\end{eqnarray}
The source terms for these equations are provided by our initial state model.

The energy-momentum source $S^{\mu}$ is obtained from all the energy-momentum depositions as
\begin{equation}\label{eq:em_source}
    S^{\mu}_{\mathrm{source}} = \sum_i p_{i}^{\mu} \, f_{\mathrm{smear}}.
\end{equation}
Here $p^{\mu}_i = (E_i,0,0,p_{z,i})$ is the energy-momentum of binary collisions events and of the participants at the local space-time point. We use a Gaussian smearing profile $f_{\mathrm{smear}}$ given by
\be
\label{eq:smearing}
    f_{\mathrm{smear}} = \frac{1}{N} \exp\left[-\frac{(x_i-x_{c})^{2}+(y_i-y_{c})^{2}}{\sigma_{\perp}^{2}}
    -\frac{(\eta_i-\eta_{c})^{2}}{\sigma_{\eta}^{2}}\right],
\ee
with
\begin{equation}
    N = \Delta\tau(2\pi)^{(3/2)}\sigma_{\perp}^{2}\tau_i\sigma_{\eta}.
\end{equation}
Here $(\tau_i,x_i,y_i,\eta_i)$ are local the space-time coordinates on the hydro grid, $\Delta\tau$ is the time-step size on the hydrodynamic solver, and the transverse smearing width $\sigma_{\perp}$ is chosen to be 0.5 fm. The energy-momentum source from the participants can be similarly obtained using Eq. (\ref{eq:em_source}), with the space-time positions of the collisions being replaced by the final positions of the participants. For the participants, the transverse smearing width $\sigma_\perp$ is kept the same as 0.5 fm while the longitudinal smearing width $\sigma_\eta$ is chosen to be 0.2.

The baryon current source $\rho_{B}$ has contributions from every participant given by
\begin{eqnarray}
    \rho_{B,\mathrm{source}} &=& \sum_{i} b_i \frac{u_{\mu}p_i^{\mu}}{p^{\tau}_i} f_{\mathrm{smear}}\nonumber\\ 
    &=& \sum_{i} b_i \left[ u^{0} + \frac{u^{3}p_{z,i}}{E_i}\right] f_{\mathrm{smear}}
\end{eqnarray}
The local fluid four-velocity is given by $u^{\mu}$ and $b_i$ is the baryonic charge.

\section{The equation of state}\label{Sec:EOS}
The equation of state used in the hydrodynamic stage of the model is taken from Ref. \cite{Albright:2014gva}. The pressure is expressed as the sum of contributions from a hadron resonance gas (HRG) model with an excluded volume correction, and a model obtained from perturbative QCD as

\begin{eqnarray}
    P(T,\mu_B) &=& (1-S(T,\mu_B)) P_{\mathrm{HRG}}(T,\mu_B)\nonumber\\
 & & + S(T,\mu_B) P_{\mathrm{pQCD}}(T,\mu_B)
\end{eqnarray}
The function $S(T,\mu_B)$ takes values between $0$ and $1$ and parameterizes the degree to which the hadronized and dehadronized phases contribute. Its functional form is given by:
\begin{equation}
S(T,\mu_B)=\exp \left[-\left( \frac{T^2}{T_0^2} +\frac{\mu_B^2}{\mu_0^2}\right)^{-\frac{r}{2}}\right]
\end{equation}
The constants $T_0 =177.12 \, \textrm{MeV}$ and $r=5$, which is restricted to integers, are  determined by fitting to lattice values at $T=0$. We take $\mu_0= 3\pi T_0$. This function is constructed to be strictly increasing in $T$ and $\mu_B$. It is smooth and infinitely differentiable so as not to introduce discontinuities which would cause a phase transition. 

The function $P_{\mathrm{HRG}}(T,\mu_B)$ is the pressure for an ideal gas of hadron resonances with an excluded volume from each particle proportional to its total energy. The proportionality constant $\epsilon_0 = 1.15$ GeV/fm$^3$ is determined from a fit to lattice values. We include all hadrons composed of u, d, and s quarks as listed by Ref. \cite{ParticleDataGroup:2020ssz}.

The function $P_{\mathrm{pQCD}}(T,\mu_B)$ is the standard EOS determined from perturbative QCD involving 3 flavors of massless quarks \cite{Vuorinen:2003fs} with slight modifications. The first is that the renormalization scale for the running coupling is given by $M=C_M \sqrt{(\pi T)^2 + (\mu_B /3)^2}$, with the constant $C_M =3.352$ determined via fit to lattice data. Second, since the running coupling depends upon $M$ through the quantity $t=\ln(M^2 / \Lambda_{\overline{MS}}^2)$, there will be a divergence at low temperatures, where the pressure should only depend on the HRG model anyway. To regulate this divergence, $t$ is replaced with $t=\ln(C_S^2 + M^2 / \Lambda_{\overline{MS}}^2)$, with the constant $C_S= 4.28$ again being determined from fitting. The $\Lambda_{\overline{MS}}$ is taken to be $290\, \textrm{MeV}$.

From $P(T,\mu_B)$, the baryon number, entropy and energy densities can be determined from
\begin{equation}
s=\left( \frac{\partial P}{\partial T} \right)_{\mu_B} \; , \quad
n_B=\left( \frac{\partial P}{\partial \mu_B} \right)_{T}
\end{equation}
\begin{equation}
\epsilon=T s + \mu_B n_B - P
\end{equation}
For the purposes of our model, $T, \mu_B, P$ and $s$ are tabulated as functions of $n_B$ and $\epsilon$ through numerical root finding.

\section{Departure Functions}\label{Sec:Departure_Functions}
To first order in a departure from equilibrium the quasiparticle distribution function for species $a$ is
\be
f_a = f_a^{\rm eq} \left( 1 + \phi_a \right)
\ee
where $f_a^{\rm eq}$ is the distribution function in thermal and chemical equilibrium.  In what follows we use a relativistic Boltzmann distribution
\be
f_a^{\rm eq} = \exp[-(E_a - \mu_a)/T]
\ee
where $\mu_a = b_a \mu_B$ and $b_a$ is the baryon number of species $a$.  To match the viscous and thermal conduction contributions to the energy-momentum tensor and baryon current the $\phi_a$ must have the form 
\ba
\phi_a &=& -A_a \partial_\rho u^\rho - B_a p_a^{\nu} D_\nu \left( \frac{\mu_B}{T} \right) \nonumber \\
&+& C_a p_a^{\mu} p_a^{\nu} 
\left( D_\mu u_\nu + D_\nu u_\mu + \twoth \Delta_{\mu\nu}\partial_\rho u^\rho \right)
\label{eq:derivingphi:phi}
\ea
The functions $A_a$, $B_a$ and $C_a$ only depend on momentum $p$ while the 4-velocity $u^{\mu}$ only depends on space-time coordinate $x$.  Detailed studies were carried out at zero baryon density \cite{Chakraborty2011} and later extended to nonzero baryon density \cite{Albright:2015fpa}.  One can express the departure from equilibrium as
\ba
f_a(E_a, T, \mu_B) &=& f_a^{\rm eq}(E_a^0, T^0, \mu_B^0) + \delta f_a \nonumber \\
&=& f_a^{\rm eq}(E_a, T^0, \mu_B^0) + \delta \tilde{f}_a
\ea
Here $E_a^0$ denotes the equilibrium single particle energy and $E_a$ the total nonequilibrium energy; it is the latter which is conserved in the particle collisions.  It is the $\delta \tilde{f}_a$ which determine the transport coefficients, not the $\delta f_a$.  They are related by 
\be
\delta f_a = \left[ 1  -   \frac{ T( \partial E_a/\partial T)_{\sigma}}{E_a - \mu_a + T \left( \partial \mu_a / \partial T \right)_{\sigma} }\right] \delta \tilde{f}_a
\ee
where $\sigma$ is the entropy per baryon.  If there are no mean fields (scalar or vector) present then $(\partial E_a/\partial T)_{\sigma} = 0$ and $\delta f_a = \delta \tilde{f}_a$.  This is the situation we assume here.

It is useful to know the contributions to the pressure, energy density, baryon density, entropy density, and heat capacity from a single species of particle. 
\ba
P_a &=& T \int d\Gamma_a f_a^{\rm eq} = \int d\Gamma_a \frac{p^2}{3 E_a} f_a^{\rm eq} \nonumber \\
\epsilon_a &=& \int d\Gamma_a E_a f_a^{\rm eq} \nonumber \\
n_{Ba} &=& b_a n_a = \frac{b_a P_a}{T} \nonumber \\
T s_a &=& \frac{1}{3T} \int d\Gamma_a p^2 f_a^{\rm eq} - \mu_B n_{Ba} \nonumber \\
\ea
where
\be
d\Gamma_a = (2s_a+1) \frac{d^3p_a}{(2\pi)^3}
\ee

\subsection{Shear Viscosity}

The shear viscosity is
\be
\eta = \frac{2}{15} \sum_a \int d\Gamma_a \frac{p^4}{E_a} f_a^{\rm eq} C_a
\label{eta}
\ee
The usual simplifying assumption when computing the departure distribution during particlization is that $C_a$ is independent of energy and of particle species \cite{degroot,Teaney:2003kp}.  Integration by parts gives
\be
C_a = \frac{\eta/w}{2T^2}
\label{Cw}
\ee
where enthalpy density is $w = T s + \mu_B n_B = \epsilon + P$.  This generalizes the oft-used formula
\be
C_a = \frac{\eta/s}{2T^3}
\ee
to nonzero baryon density.

In the relaxation time approximation
\be
C_a = \frac{\tau_a(E_a)}{2TE_a}
\ee
Numerical results from the linear $\sigma$ model suggest that $\tau_a(E_a) \propto E_a$ \cite{Chakraborty2017}.  This makes physical sense since higher momentum or higher mass particles should take longer to reach kinetic equilibrium.  If this is the case, and if one takes $\tau_a(E_a) = \tau' E_a$ with $\tau'$ a constant, then 
\ba
C_a &=& \frac{\tau'}{2T} \nonumber \\
\tau' &=& \frac{\eta}{2Tw}
\ea
where $\tau'$ has units of 1/energy$^2$ and is independent of species.  This may be used to inform the thermal conductivity and bulk viscosity.

\subsection{Thermal Conductivity}

The formula for the thermal conductivity associated with the baryon current is
\be
\lambda = \frac{1}{3} \left( \frac{w}{n_B T} \right)^2 \sum_a b_a \int d\Gamma_a \frac{p^2}{E_a} f_a^{\rm eq} B_a \,.
\label{lambda}
\ee
Due to energy-momentum conservation, if we have a particular solution $B_a^{\rm par}$ to the integro-differential equation arising from the Boltzmann equation we can generate another solution as $B_a = B_a^{\rm par} - b$, where $b$ is a constant independent of particle species $a$.  This freedom is resolved by the Landau-Lifshitz condition of fit which requires that $\delta T^{0j} = 0$ in the local rest frame.  The result is that
\be
b  = \frac{1}{3Tw} \sum_a \int d\Gamma_a p^2  f_a^{\rm eq} B_a^{\rm par}
\ee
where
\be
3 T w = \sum_a d \Gamma_a p^2 f_a^{\rm eq}
\ee
Substitution into expression (\ref{lambda}) gives 
\be
\lambda = \frac{1}{3} \left( \frac{w}{n_B T} \right)^2 \sum_a \int d\Gamma_a \frac{p^2}{E_a} 
\left( b_a - \frac{n_B E_a}{w} \right)  f_a^{\rm eq} B_a^{\rm par}
\label{lambdaLL}
\ee

The simplest approximation is to take $B_a^{\rm par}$ = constant, as in the case of the shear viscosity.  However, it is easily shown that this results in $\lambda = 0$.    Alternatively, with this approximation one gets $b = B_a^{\rm par}$, and so $B_a = 0$ and again $\lambda = 0$.   

In the relaxation time approximation
\be
B_a^{\rm par} = \frac{\tau_a(E_a)}{E_a} \left( b_a - \frac{n_B}{w} E_a \right)
\label{eq:RTA:Bpar}
\ee

To be consistent with expression (\ref{Cw}) one uses $\tau_a(E_a) = \tau' E_a$.  Then
\be
b = \frac{\tau'}{3Tw} \left[ \sum_a b_a \int d\Gamma_a p^2 f_a^{\rm eq} - \frac{n_B}{w} \sum_a \int d\Gamma_a p^2 E_a f_a^{\rm eq} \right]
\ee
After some manipulation of integrals and using thermodynamic identities this can be written as
\be
b = \frac{\tau' T}{w^2} \left[ s \sum_a b_a w_a - n_B w - n_B T \frac{\partial w}{\partial T} \right]
\label{bw}
\ee
and then as
\ba
b &=& \frac{\tau' T}{w^2} \Big[ T s (T \chi_{T\mu} + \mu_B \chi_{\mu\mu} ) \nonumber \\
&-&  T n_B (T \chi_{TT} + \mu_B \chi_{T\mu} ) - n_B w \Big]
\ea
Here the susceptibilities are
\be
\chi_{xy} = \frac{\partial^2 P(T,\mu)}{\partial x \partial y}
\ee

Finally
\be
B_a = \tau' \left( b_a - \frac{n_B}{w} E_a \right) - b(T,\mu_B)
\ee
Unfortunately the expression for $B_a$ is not a simple pocket formula as it is for $C_a$.

Using Boltzmann statistics, the integrals we need are
\ba
&& \sum_a b_a^2 \int d\Gamma_a  \frac{p^2}{E_a} f_a^{\rm eq} = 3 T^2 \chi_{\mu\mu} \nonumber \\
&& \sum_a b_a \int d\Gamma_a p^2 f_a^{\rm eq} = 3 T^2 ( T \chi_{T\mu} +  \mu_B \chi_{\mu\mu} + n_B ) \nonumber \\
&& \sum_a \int d\Gamma_a p^2 E_a f_a^{\rm eq} = 6 T^2 w \nonumber \\
&& + 3 T^2 ( T^2 \chi_{TT} + 2 T \mu_B \chi_{T\mu} +  \mu_B^2  \chi_{\mu\mu} )
\ea
which results in
\ba
\lambda &=& \tau' T^2 \left[ \left( \frac{s}{n_B}\right)^2 \chi_{\mu\mu} - 2 \left( \frac{s}{n_B}\right) \chi_{T\mu} + \chi_{TT} \right] \nonumber \\
&=& \tau' \left( \frac{T}{n_B} \right)^2 w \det \chi v_{\sigma}^2 = 
\frac{\eta T}{2} \frac{\det \chi}{n_B^2} v_{\sigma}^2
\ea
where $v_{\sigma}$ is the speed of sound (see below) and
\be
\det \chi = \chi_{TT} \chi_{\mu\mu} - \chi^2_{T\mu}
\ee

\subsection{Bulk Viscosity}

The bulk viscosity is
\be
\zeta = \frac{1}{3} \sum_a \int d\Gamma_a \frac{p^2}{E_a} f_a^{\rm eq} A_a \,.
\label{zeta}
\ee
In the relaxation time approximation the particular solution is
\be
A_a^{\rm par} = \frac{\tau_a}{3T} \left\{ \frac{p^2}{E_a} -3 \left[ v_n^2 E_a + (v_s^2 - v_n^2) \mu_a \right] \right\}
\ee
where, for brevity of notation, we have defined
\ba
v_n^2 &=& \left(\frac{\partial P}{\partial \epsilon}\right)_n = 
\frac{s \chi_{\mu\mu} - n_B \chi_{\mu T}}{T \det \chi}
\nonumber \\
v_s^2 &=& \left(\frac{\partial P}{\partial \epsilon}\right)_s = 
\frac{n_B \chi_{TT} - s \chi_{\mu T}}{\mu_B \det \chi}
\nonumber \\
v_{\sigma}^2 &=& \left(\frac{\partial P}{\partial \epsilon}\right)_{\sigma} =
\frac{v_n^2 T s + v_s^2 \mu_B n_B}{w}
\label{speeds}
\ea
Of course waves do not physically propagate at constant $n$ or $s$, only at constant $\sigma$.

In order to satisfy the Landau-Lifshitz condition of fit we must allow for the functional form
\be
A_a = A_a^{\rm par} - a_E E_a - a_B b_a
\ee
Here
\ba
a_E &=& \frac{X_B Z_B - Y_B Z_E}{Y_E X_B - X_E Y_B} \nonumber \\
a_B &=& \frac{Y_E Z_E - X_E Z_B}{Y_E X_B - X_E Y_B}
\ea
where
\ba
X_E &=& T(T^2 \chi_{TT} + 2 \mu_B T \chi_{T \mu} + \mu_B^2 \chi_{\mu\mu})  \nonumber \\
X_B &=& T( T \chi_{T \mu} + \mu_B \chi_{\mu\mu}) \nonumber \\
Y_E &=& T( T \chi_{T \mu} + \mu_B \chi_{\mu\mu}) \nonumber \\
Y_B &=& T \chi_{\mu\mu}
\ea
and
\ba
Z_E &=& \sum_a \int d\Gamma_a E_a  A_a^{\rm par} f_a^{\rm eq} \nonumber \\
Z_B &=& \sum_a b_a \int d\Gamma_a  A_a^{\rm par} f_a^{\rm eq}
\ea

As in the case of thermal conductivity, one should take $\tau_a = \tau' E_a$ to be consistent with the standard result for the shear viscosity.  This leads to complicated formulas for $Z_E$ and $Z_B$.  To evaluate the lengthy integrals it is useful to have the identities
\ba
T \frac{\partial}{\partial \mu_B} f_a^{\rm eq} &=& b_a  f_a^{\rm eq} \nonumber \\
T^2 \frac{\partial}{\partial T} f_a^{\rm eq} &=& (E_a - \mu_a) f_a^{\rm eq} \nonumber \\
T \left( T \frac{\partial}{\partial T} + \mu_B \frac{\partial}{\partial \mu_B} \right)  f_a^{\rm eq} &=& E_a  f_a^{\rm eq}
\ea
which are valid for Boltzmann statistics.  The results are
\ba
&& Z_B/T \tau' = n_B + (1-v_n^2 - v_s^2 ) \mu_B \chi_{\mu\mu} + (1-2v_n^2) T \chi_{T\mu} \nonumber \\
&& - (v_n^2 + v_s^2) T \mu_B \chi_{T\mu\mu} -  v_n^2 T^2 \chi_{TT\mu} -  v_s^2 \mu_B^2 \chi_{\mu\mu\mu}
\ea
and
\ba
&& Z_E/T \tau' = 2w + (1-3v_n^2) T^2 \chi_{TT} \nonumber \\
&& + 2 (1-2v_n^2 - v_s^2 ) T \mu_B \chi_{T\mu}
+ (1-v_n^2 - v_s^2 ) \mu_B^2 \chi_{\mu\mu} \nonumber \\
&& - v_n^2 T^3 \chi_{TTT} - (2v_n^2 + v_s^2) T^2 \mu_B \chi_{TT\mu} \nonumber \\
&& - (v_n^2 + 2v_s^2) T \mu_B^2  \chi_{T\mu\mu} - v_s^2 \mu_B^3 \chi_{\mu\mu\mu}
\ea

The bulk viscosity can be expressed in terms of the equation of state and susceptibilities, but the expression is very long and not displayed here.

\subsection{Relaxation time for baryon current}

Now the question is how to choose the relaxation time for the baryon current $\tau_B$.  This is the time constant that appears in the Cattaneo equation \cite{Young,Plumberg}.  Properly speaking the baryon diffusion constant is $D_B$ and has units of length (in natural units with $c = 1$).  
\be
D_B = \frac{\lambda T}{\chi_{\mu\mu}} \left( \frac{n_B}{w}\right)^2 = \frac{\kappa_B}{T \chi_{\mu\mu}}
\ee
Collective baryon fluctuations travel with speed $v_B^2 = D_B/\tau_B$ \cite{Young,Plumberg}.  This means that $\tau_B > D_B$ in order that signals not travel faster than the speed of light.  Using the above formulas for $\lambda$ and $\tau'$ we have
\be\label{eq:baryon_diffusion}
D_B = \frac{\eta}{2} \left( \frac{T}{w}\right)^2 \frac{\det \chi}{\chi_{\mu\mu}} v_{\sigma}^2
\ee
This expression is simplified if all the particles are assumed to have a constant relaxation time \cite{McGill1}. The two expressions are compared in appendix \ref{App:compare_conductivity}.

The two speeds $v_B$ and $v_{\sigma}$ cannot be too different, so we choose them equal to obtain
\be\label{eq:baryon_relaxation}
\tau_B = \frac{\eta}{2} \left( \frac{T}{w}\right)^2 \frac{\det \chi}{\chi_{\mu\mu}}
\ee

\begin{figure*}[]
    \centering
    \includegraphics[scale=1.5]{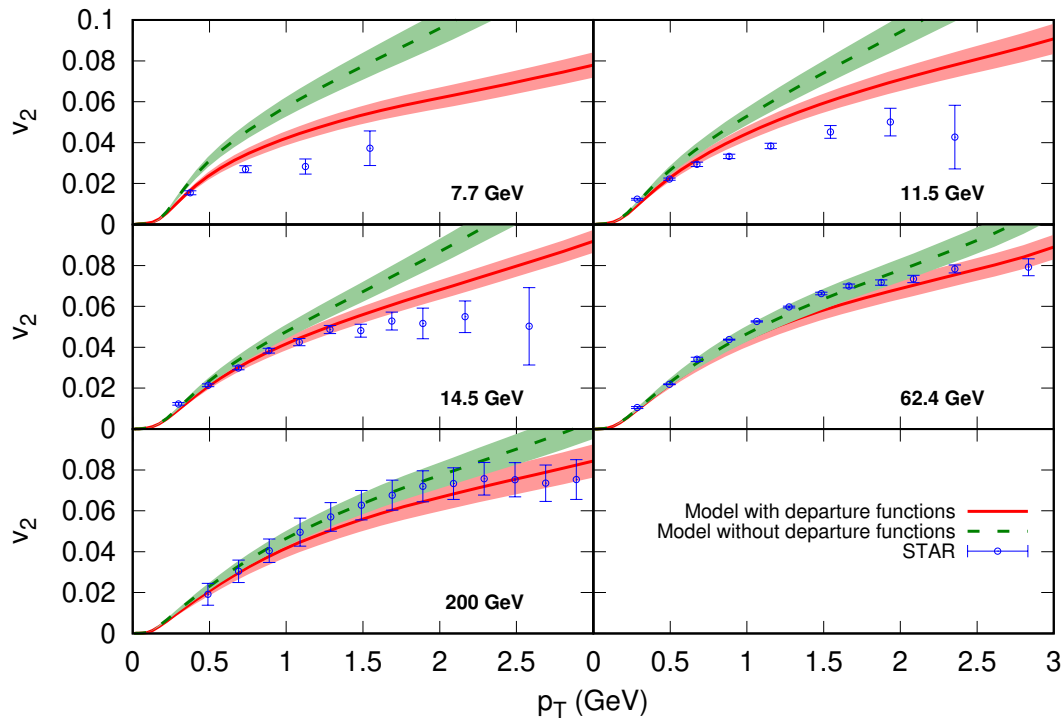}
    \caption{The $v_2$ for $\pi^{+}$ for the centrality bin $0-10\%$. Shaded areas represent statistical uncertainties.  Experimental data are from \cite{PHENIX:2015tbb,STAR:2015rxv}.}
    \label{fig:v_2_pi}
\end{figure*}

\begin{figure*}[]
    \centering
    \includegraphics[scale=1.5]{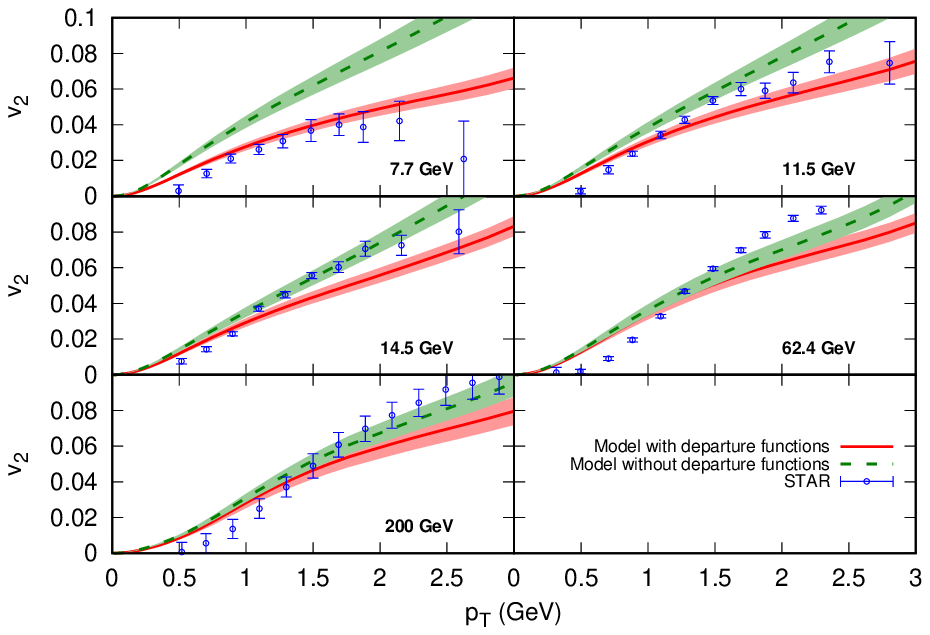}
    \caption{The $v_2$ for proton for the centrality bin $0-10\%$. Shaded areas represent statistical uncertainties.  Experimental data are from \cite{PHENIX:2015tbb,STAR:2015rxv}.}
    \label{fig:v_2_proton}
\end{figure*}

\section{Results}
\label{Sec:Results}

Our goal is to conduct realistic heavy-ion collision simulations at finite baryon chemical potential with a LEXUS-inspired initial-state model, a crossover EOS without a critical point, and departure functions derived using quasiparticle theory at non-zero baryon chemical potential. In many regards, it is a straightforward modeling of the heavy ion simulations to test our physics understanding of the initial state at finite baryon chemical potential and incorporating departure functions that are specifically calculated for non-zero baryon chemical potential. In this section we compare our simulations with experimental data from RHIC. We calculate the single particle spectra and the flow harmonics.

Our LEXUS-inspired model provides the initial conditions for MUSIC.  The produced matter is then  hydrodynamically evolved until freezeout which is assumed to occur on the constant energy density surface 0.205 GeV/fm$^3$. We use the baryon diffusion constant and baryon relaxation times given in equations (\ref{eq:baryon_diffusion}) and (\ref{eq:baryon_relaxation}). The particle spectra are computed using the Cooper-Frye procedure and our new departure functions.  The Cooper-Frye procedure is performed using mode 3 in MUSIC and the obtained particle distributions undergo resonance decays using mode 4 in MUSIC. All hadronic resonances with mass less than 1.8 GeV in the PDG table \cite{ParticleDataGroup:2020ssz} were included in our study. No hadronic after-burner was employed in this work. Effects of an hadronic afterburner is left for future investigation.

Impact parameters are randomly sampled between $0$ and $20$ fm. The centrality is then determined by categorizing the $5\%$ of events with the highest total energy as the $0-5\%$ central events, the next $5\%$ events in terms of total energy as the $5-10\%$ central events, and so on. This process is close to what is done in experiments, which bin events on charged particle multiplicity. Thus, the total energy in the initial state is used as a proxy for the final multiplicity.

\subsection{Hydrodynamic flow}

One of the key signatures of flow in heavy ion collisions is the second harmonic coefficient of particle multiplicity $v_2$.
\begin{align}
    \frac{dN}{p_Tdp_Tdyd\phi} = \frac{dN}{2\pi p_T dp_T dy} \left[ 1 + \sum_n 2v_n(y,p_T)\right.\nonumber\\
    \times \left. \; \cos (n\phi-n\Psi_n(p_T))\vphantom{\sum_n 2v_n(y,p_T)}\right]
\end{align}
The event-plane angles $\Psi_n(p_T)$ are determined event by event and given as
\begin{equation}
    \Psi_n(p_T) = \left(\tan^{-1}\frac{\sum_i\sin(n\phi_i)}{\sum_i\cos(n\phi_i)}\right)/n.
\end{equation}

We compute the $v_2$ flow harmonic as a function of $p_T$. The shear viscosity to entropy ratio $\eta/s$ that we use is a constant for a given collision energy and not T and $\mu_B$ dependent. The ratio $\eta/s$ was adjusted to match the $p_T$-differential $v_2$ data for various collision energies in the $0-10\%$ centrality class as a function of $p_T$, when departure functions are included. Calculations without departure functions are done with same value of $\eta/s$. Table \ref{table:eta_s} lists the $\eta/s$ for various collision energies.

\begin{table}[h!]
\centering
\begin{tabular}{| c| c| }
 \hline
 Collision energy & $\eta/s$ \\
 \hline 
 200 GeV & 0.08 \\  
 62.4 GeV & 0.08 \\
 14.5 GeV & 0.08 \\
 11.5 GeV & 0.08 \\
 7.7 GeV & 0.14\\
 \hline
\end{tabular}
\caption{$\eta/s$ for various collision energies.}
\label{table:eta_s}
\end{table}

Figure \ref{fig:v_2_pi} shows the $p_T$-differential $v_2$ for $\pi^{+}$ at midrapidity for the above mentioned energies compared to STAR data. Theoretical calculations are shown with the statistical error band. There is good agreement between the simulation and the experimental data. The agreement with STAR data for protons for the $0-10\%$ centrality class is also reasonable as shown in Fig. \ref{fig:v_2_proton}. 

One can see that the agreement with the data ceases to be very good beyond $p_T$ of about 1.5 GeV/c. That is to be expected because hydrodynamics is a long-wavelength theory and is not particularly effective for large $p_T$.

\begin{figure*}
    \centering
    \includegraphics[scale=1.5]{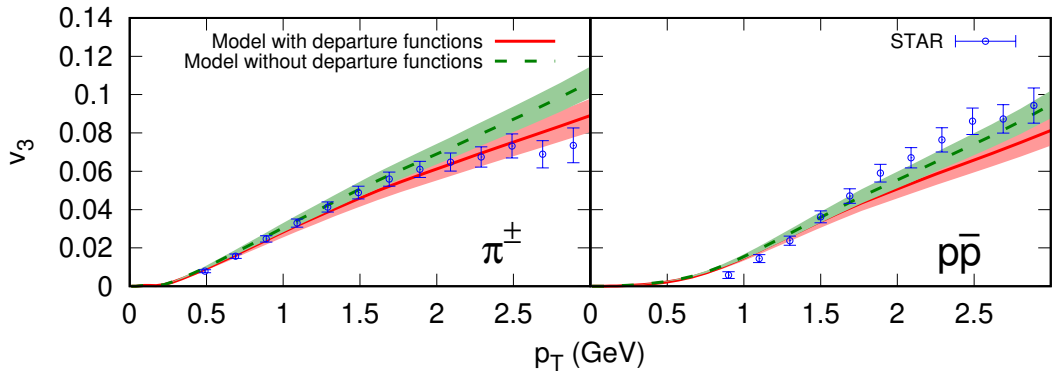}
    \caption{The $v_3$ for $\sqrt{s_{NN}} = 200$ GeV Au+Au collisions for centrality $0-10\%$. Shaded areas represent statistical uncertainties.  Experimental data are from from \cite{PHENIX:2014uik}.}
    \label{fig:v3_200}
\end{figure*}

\begin{figure*}
    \centering
    \includegraphics[scale=1.5]{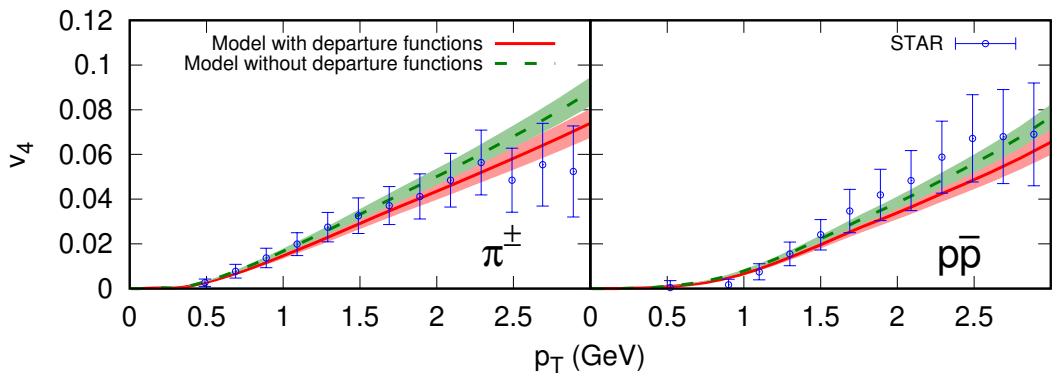}
    \caption{The $v_4$ for $\sqrt{s_{NN}} = 200$ GeV Au+Au collisions for centrality $0-10\%$. Shaded areas represent statistical uncertainties.  Experimental data are from \cite{PHENIX:2014uik}.}
    \label{fig:v4_200}
\end{figure*}

The $v_2$ is mainly driven by the initial geometry of the system. Higher harmonics are more sensitive to small-scale initial fluctuations. The $v_3$ and $v_4$ measurements for 200 GeV collisions are available and are compared to the model predictions in Figs. \ref{fig:v3_200} and \ref{fig:v4_200}, respectively. Our results are in good agreement with the experimental data for $v_3$ and $v_4$ as well.

Departure functions have a sizeable effect on $p_T$ differential $v_n$s. This effect is more pronounced as we go to lower collision energies where baryon densities are higher. Inclusion of departure functions requires lower values of $\eta/s$ to explain the data.

\begin{figure*}[]
  \centering
  \includegraphics[scale=1.5]{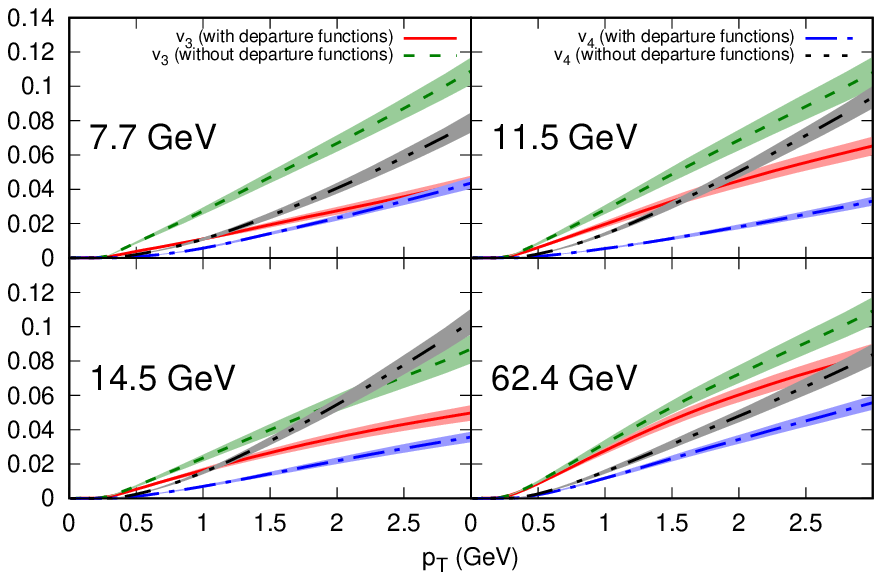}
  \caption{Predictions for $v_3$ and $v_4$ for $\pi^{+}$ for the centrality bin $0-10\%$.  Shaded areas represent statistical uncertainties.}
  \label{fig:v_34_pip_prediction}
\end{figure*}

We present our predictions for the $\pi^{+}$ $v_3$ and $v_4$ for BES energies in Fig. \ref{fig:v_34_pip_prediction}. These should be compared to BES measurements when the data becomes available \cite{Parfenov:2020fuo}.

\begin{figure*}[ht!]
  \centering
  \includegraphics[scale=0.5]{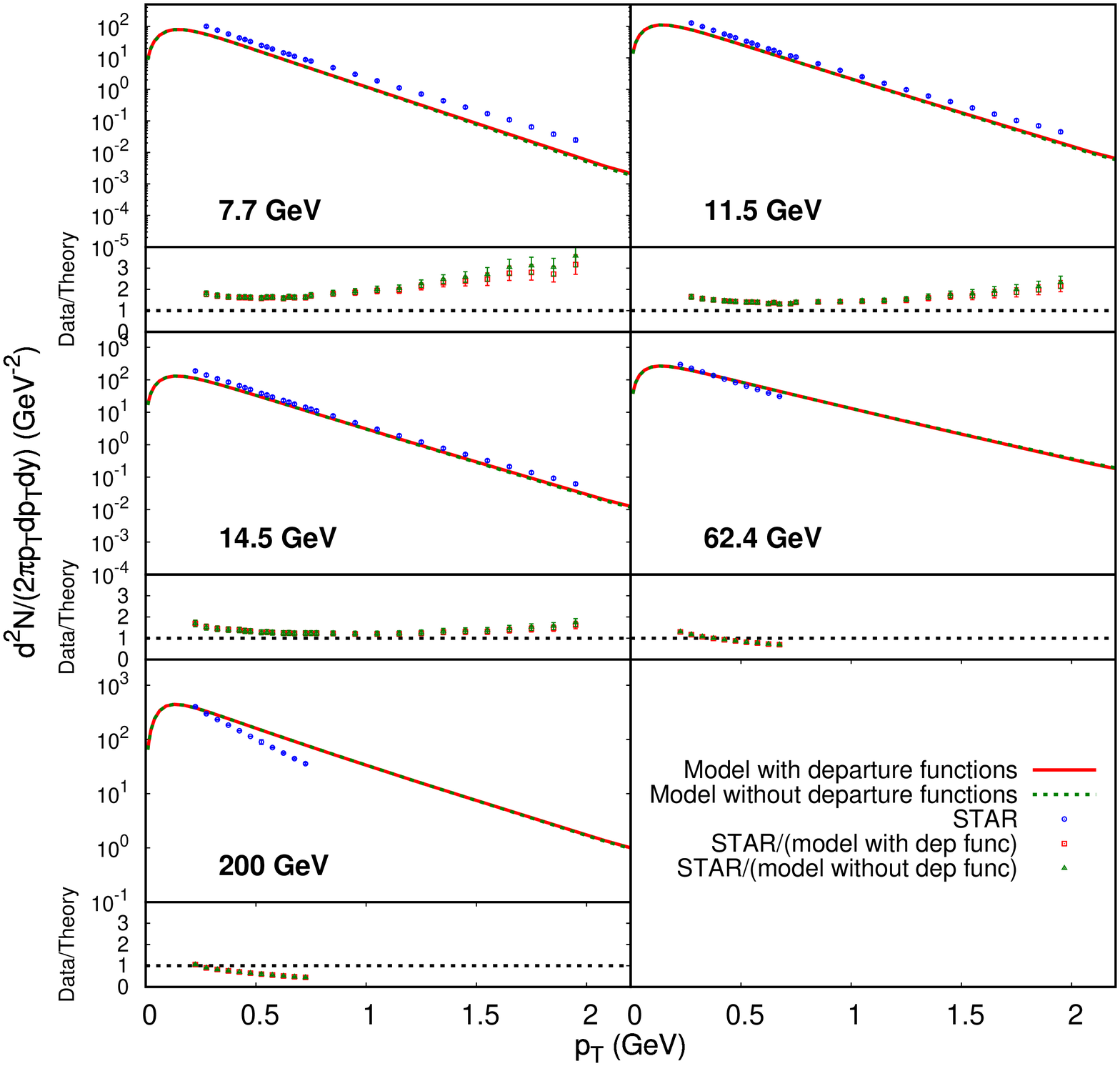}
  \caption{Pion transverse momentum distributions for Au+Au collision for the centrality bin $0-5\%$. Experimental data are from \cite{STAR:2008med,STAR:2017sal,STAR:2019vcp}.}
  \label{fig:pion_multiplicity}
\end{figure*}

\begin{figure*}[ht!]
  \centering
  \includegraphics[scale=0.5]{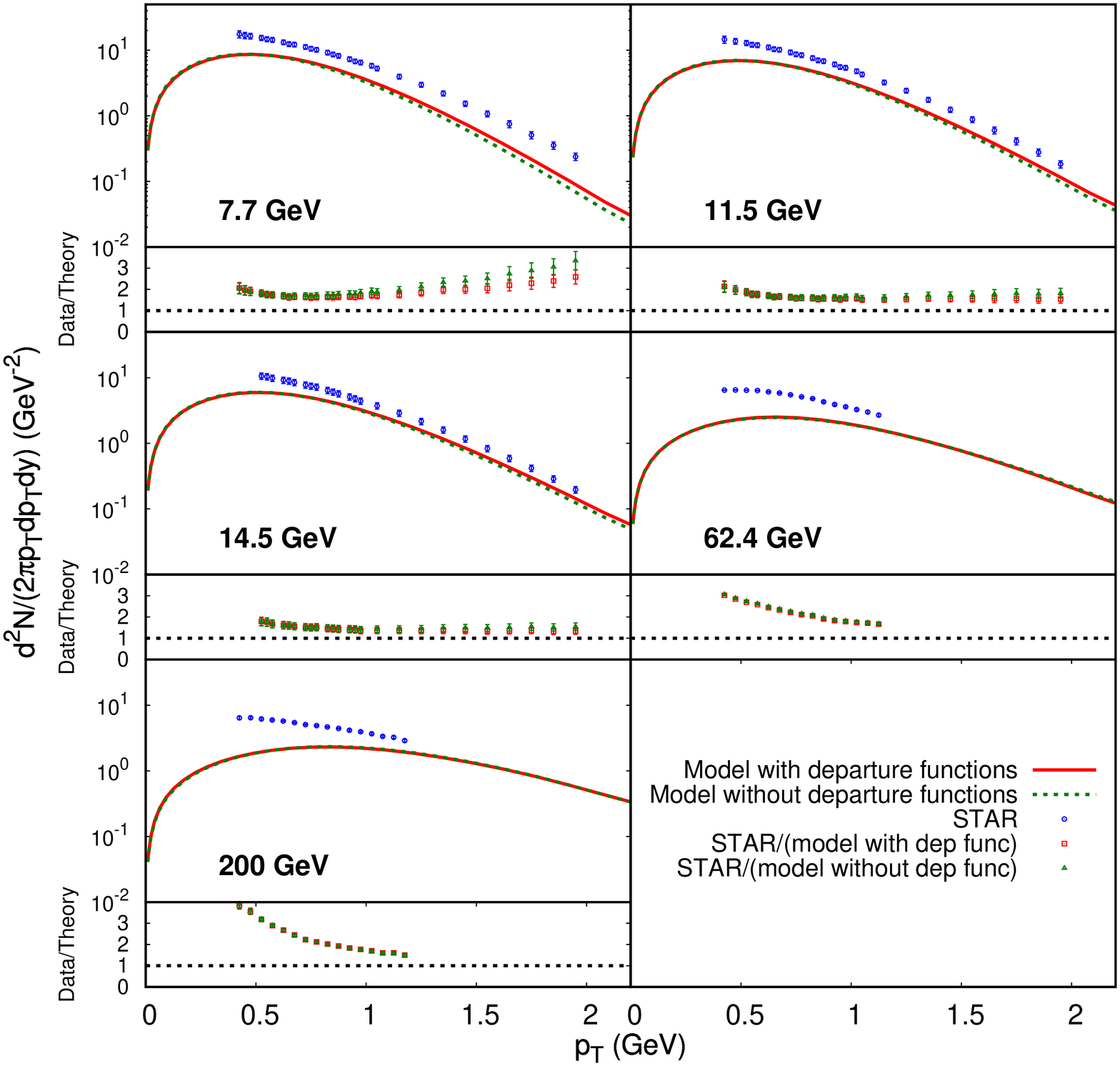}
  \caption{Proton transverse momentum distributions for Au+Au for the centrality bin $0-5\%$. Experimental data are from \cite{STAR:2008med,STAR:2017sal,STAR:2019vcp}.}
  \label{fig:proton_multiplicity}
\end{figure*}

\subsection{Effect of the departure functions on hadron rapidity and transverse momentum distributions}

We are working under the assumption of an overlapping period of applicability between hydrodynamic and kinetic theories which lets us match the energy-momentum tensor and baryon current from fluid dynamics to particle distributions in the kinetic theory. We calculate $p_T$ spectra for pions and protons with and without the use of departure functions. The results are compared to experimental data. These comparisons are shown in Figs.\ref{fig:pion_multiplicity} and \ref{fig:proton_multiplicity} for five collision energies. There is an enhancement when we include the $\delta f$ terms, which is to be expected, since there is an increase in the hadrons being sampled because of the out-of-equilibrium corrections. Corrections due to the $\delta f$ terms are more prominent for the proton yield than for the pion yield. That makes sense because the out-of-equilibrium corrections depend on the relaxation times. We have taken the relaxation times to be proportional to the mass of the hadronic species; hence, the contribution of the $\delta f$ terms to the particle multiplicities are greater in the proton yields compared to the pion yields.

 In general, we expect our LEXUS based initial state model to be more accurate for low-energy heavy-ion collisions compared to high-energy heavy-ion collisions. The reason is that we are treating nucleus-nucleus collisions as a sequence of nucleon-nucleon collisions. One would surmise that as one goes to higher and higher collision energies, there will be additional physics from partonic degrees of freedom that are missing from a LEXUS based model. Our aim is to give a 3D model for lower-energy collisions, so we 
 should not be too concerned about discrepancies at higher-energy collisions. 
 
There are various other factors that might contribute to the discrepancies with the data. We are not using an hadronic afterburner in this work; this is because we want to focus on the effects of the departure functions $\delta f$. Using an hadronic afterburner will require particle sampling and will increase the computational cost to achieve similar statistics. The hadronic rescatterings among light mesons and baryons largely blue-shift the distributions and shift them to higher $p_T$. This is the pion wind effect that pushes the heavier particles to the high $p_T$ region \cite{Ryu:2015vwa}. The effect should be more pronounced in the case of protons than pions.  If one compares Figs. \ref{fig:pion_multiplicity} and \ref{fig:proton_multiplicity}, the difference with data is greater for protons. Hence, the addition of hadronic rescatterings will have the desired effect. The net proton rapidity distribution is expected to be widened by scatterings with other hadrons.

We have not included contributions from weak decays of baryons which STAR has included. This will also enhance the proton multiplicity in our results.

We do not consider the possibility of transverse flow in the initial state. The presence of collective flow in the initial state might necessitate increasing the $\eta/s$ at lower energies. This would lead to more entropy production, which will enhance the multiplicities and bring them closer to the experimental data.

We are not considering bulk viscosity for the sake of simplicity. The presence of bulk viscosity would cause an increase in the multiplicities of all the hadronic species. The inclusion of bulk viscosity will involve more tuning of this work and is deferred to future investigations.

This paper includes the effects of baryon number but not electric charge. This calculation assumes that the ratio of electric charge to baryon number is one-half. Consequently, if it were not for the small difference in proton and neutron masses, the Cooper-Frye procedure would produce the same exact distribution for the two particles. As this mass difference is very small, the proton to neutron ratio is almost 1 before the resonance decays. In reality, before collisions, gold nuclei have more neutrons than protons. During the collision, there is a net conversion of neutrons to protons accompanied by more negatively charged pions than positively charged ones, but the proton to neutron ratio will be less than one. So, conserving both baryon number and electric charge requires that the proton to neutron ratio will be somewhere between $Z/A$ and 1, though in this calculation it is almost 1. This was studied long ago \cite{Kapusta:1977ce}. Doing this better will require the use of chemical potentials for electric charge and strangeness in addition to baryon number, which is left for a future investigation.

We want to emphasize that, although there is some physics missing in our initial state, our initial-state model is an extrapolation of nucleon-nucleon collision data. There is no additional normalization factor that has been introduced to match the predicted multiplicities of hadrons with the experimental data. There is a common practice of introducing normalization factors in initial-state models to match experimental data which represent unknown physics in the initial-state model, which is not necessary in this model. For better or worse,it is absolutely normalized.

\begin{figure*}[ht!]
  \centering
  \includegraphics[scale=0.5]{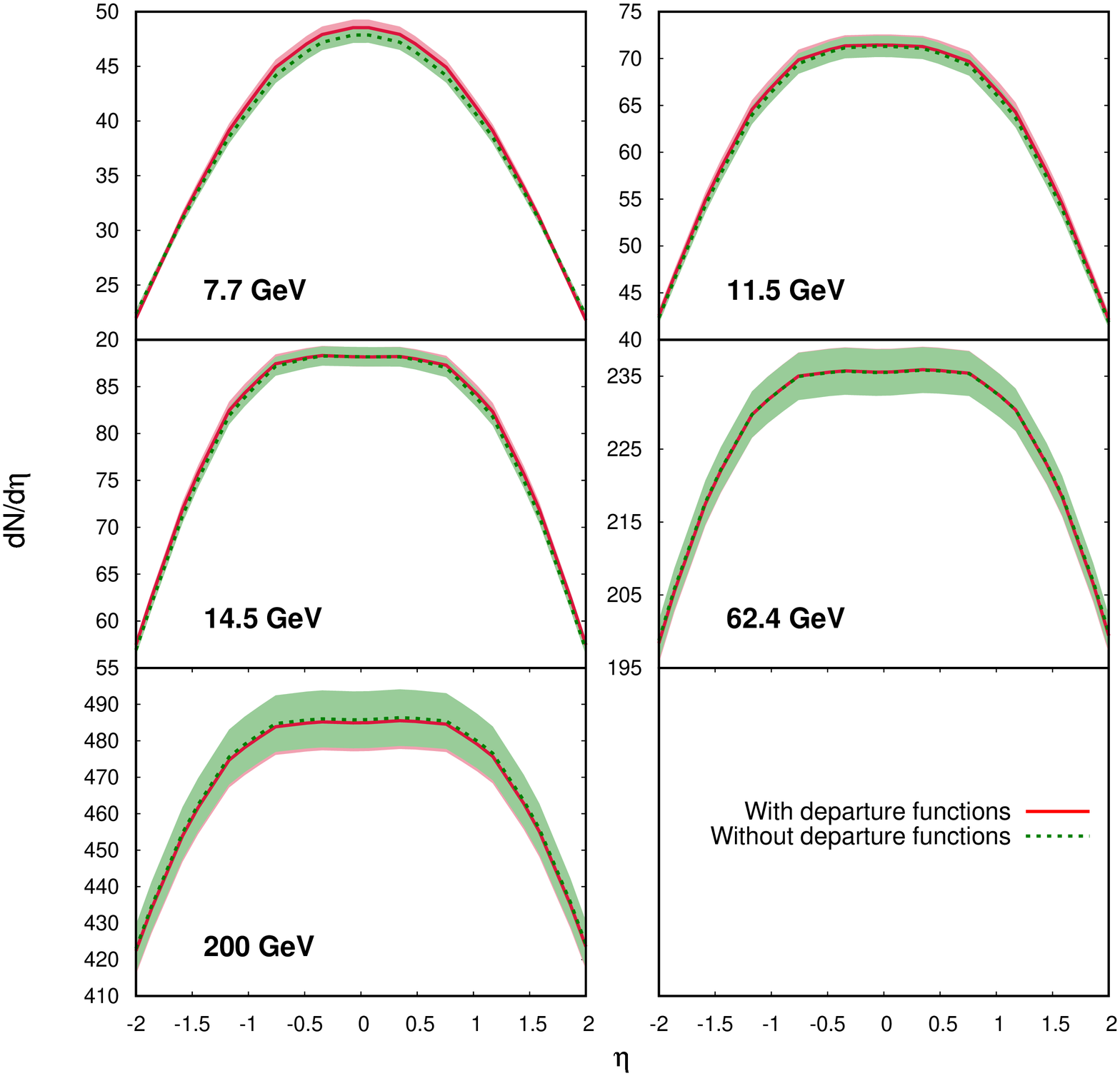}
  \caption{Pion pseudorapidity distributions for Au+Au for the centrality bin $0-5\%$. Shaded areas represent statistical uncertainties.}
  \label{fig:pion_multiplicity_eta}
\end{figure*}

\begin{figure*}[ht!]
  \centering
  \includegraphics[scale=0.5]{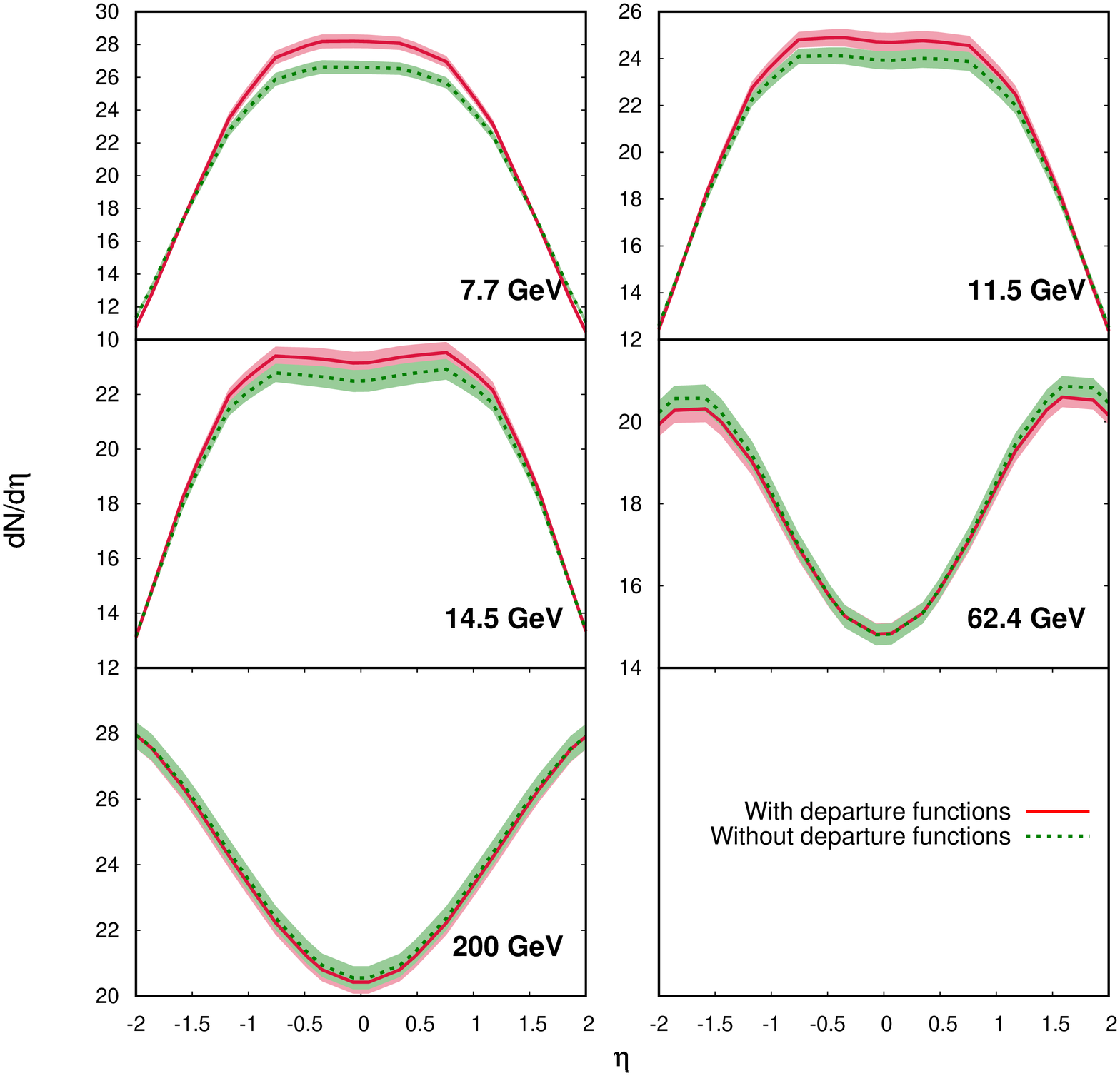}
  \caption{Proton pseudorapidity distributions for Au+Au for the centrality bin $0-5\%$.  Shaded areas represent statistical uncertainties.}
  \label{fig:proton_multiplicity_eta}
\end{figure*}

Let us now turn our attention to the rapidity distributions. The invariant yields of pions and protons as functions of pseudorapidity are plotted in Figs. \ref{fig:pion_multiplicity_eta} and \ref{fig:proton_multiplicity_eta}. Comparisons have been made to show how the distributions differ with and without the inclusion of departure functions. How they vary with pseudorapidity is more relevant for low-energy collisions where the Bjorken model is no longer a good approximation. Detailed experimental data is not yet available. If one compares the $\pi^{+}$ pseudorapidity distribution at 7.7 GeV and at 200 GeV, it is flatter at higher energies, which is a consequence of the initial binary collision energy depositions being flatter for higher energies. This is another affirmation that the Bjorken model is a good model at higher energies.

The influence of departure functions is more pronounced for proton distributions, as is evident in Fig. \ref{fig:proton_multiplicity_eta}. The consequences of the initial space-time conditions for participants are shown in Figs. \ref{fig:200GeV_gaussians_remnants} and \ref{fig:11pt5GeV_gaussians_remnants}. The separation of participant space-time deposition at higher collision energies leads to separate baryon pseudorapidity peaks present at higher energies, whereas a more uniform space-time deposition of participants at lower collision energies leads to the broader pseudorapidity distribution of protons at lower energies.

\section{Conclusions}
\label{Sec:Conclusions}

The BES program at RHIC has conducted Au+Au heavy ion collisions at center-of-momentum collision energies from 7.7 GeV to 200 GeV with the estimated baryon chemical potential ranging from 20 MeV to 420 MeV \cite{Li:2018ini}. In order to understand the experimental results and in order to `detect' the presence of the QCD critical point, we need high precision simulations at lower collision energies. This work is a baseline simulation to such studies. 

We used an initial state inspired by LEXUS.  The input is the measured binary nucleon collision cross-section. We employ dynamical initialization of the hydrodynamic solver. The hydrodynamic solver used in this paper is MUSIC \cite{Schenke:2010nt}. We use departure functions calculated at finite baryon chemical potential within the relaxation time approximation. The EOS used is a crossover equation of state without a critical point. 

We compared the transverse momentum dependent flow coefficients $v_2$ and the single particle transverse momentum distributions with the STAR data. Although we find reasonable agreement with $v_2$ for five collision energies between 7.7 GeV and 200 GeV, our model underestimates the transverse momentum distributions with respect to experimental data. We discuss the possible sources of these discrepancies. We believe that including an hadronic afterburner, and accounting for a non-zero bulk viscosity, considering initial flow in the initial state
, the hadronic multiplicities predicted with our framework will have a reasonable agreement with experimental data.

Obviously, a future direction is to incorporate hadronic scatterings and a bulk viscosity. It will be interesting to see the application of our initial state model to asymmetric collision systems like Cu+Au, 3He+Au and d+Au and see how the boost invariance is broken in this framework. One should also investigate how to include flow in the initial state and try to probe the rapidity dynamics at lower energies. We also present our predictions for higher flow harmonics which can be compared to the data when they become available \cite{Parfenov:2020fuo}. Of course, we as a community also need to find a way to simulate hydrodynamics across a critical point and across a first order phase transition, and only then we will have a full physics understanding of heavy ion collisions at finite baryon densities. We will report on progress in this direction in forthcoming publications.

\section*{Acknowledgments}
%%%%

This work was supported by the U.S. DOE Grant No. DE-FG02-87ER40328.

%%%%
\appendix
%%%%

\section{Energy loss distribution in 3D initial state}\label{App:lexus_comparison}

The normalized collision kernel in LEXUS is given in Eq. (\ref{eq:lexus_kernel}). The probability of two nucleons with rapidities $y'_p$ and $y'_T$ colliding to give nucleons with rapidities $y_P$ and $y_T$
\begin{eqnarray}
P(y'_P + y'_T \rightarrow y_P + y_T) &\propto& \cosh(y_P - y'_T)\cosh(y'_P - y_T) \nonumber \\ &=& \frac{1}{2} (\cosh(y_P-y'_T+y'_P-y_T) \nonumber \\
& & + \cosh(y_P-y'_T-y'_P+y_T))\nonumber \\
\end{eqnarray}
Recall that
\begin{eqnarray}
    y^{\text{total}}_{\text{rest-frame}} &=& y'_P - y'_T \\
    y_{\text{loss}} &=& y'_P - y'_T - (y_P - y_T)
\end{eqnarray}
Also, in pair rest frame, $y'_P + y'_T = 0$ and as both the nucleons lose $y_{\text{loss}}/2$ units in rapidity, $y_P+y_T = 0$. So,
\begin{eqnarray}
P(y_{\text{loss}})&\propto&\cosh(y_P - y'_T)\cosh(y'_P - y_T) \nonumber \\
&=& \frac{1}{2} \left(\cosh(2y^{\text{total}}_{\text{rest-frame}} - y_{\text{loss}}) + 1\right) \nonumber\\
&\approx& \frac{1}{2} \cosh(2y^{\text{total}}_{\text{rest-frame}} - y_{\text{loss}})
\end{eqnarray}

Normalizing this gives us the distribution in Eq. (\ref{eq:yloss_distribution}).

\section{Comparing Thermal Conductivities}\label{App:compare_conductivity}

Reference \cite{McGill1} uses the baryon diffusion constant
\be
\kappa_B = \oneth \tau n_B \left[ \coth\left(\frac{\mu_B}{T}\right) - \frac{3 T n_B}{w} \right]
\label{kappaB}
\ee
where $\tau$ is an energy independent relaxation time common to all particles.  For numerical studies they take $\tau = C_B/T$ and vary the dimensionless parameter $C_B$.  Where does this expression for $\kappa_B$  come from and how does it compare to ours?

First note the relationship
\be
\kappa_B = \left( \frac{T n_B}{w}\right)^2 \lambda
\ee
Expression (\ref{kappaB}) begins with quantum statistics.  In that case \cite{Albright:2015fpa}
\ba
\lambda &=& \frac{1}{3} \left( \frac{w}{n_B T} \right)^2 \sum_a \int d\Gamma_a \frac{p^2}{E_a} \frac{\tau_a(E_a)}{E_a} \nonumber \\
&\times& \left( b_a - \frac{n_B}{w} E_a \right)^2 f_a^{\rm eq} (1 - f_a^{\rm eq})
\ea
This expression already enforces the condition of fit.  When the relaxation time is the same constant for all particles the particular solution Eq. (\ref{eq:RTA:Bpar}) automatically satisfies the condition of fit without any need for a nonzero additive constant $b$ even with quantum statistics.  To see that rewrite 
\ba
&& \lambda = \frac{\tau}{3} \left( \frac{w}{n_B T} \right)^2 \sum_a b_a \int d\Gamma_a \frac{p^2}{E_a^2}
\left( b_a - \frac{n_B}{w} E_a \right) f_a^{\rm eq} (1 - f_a^{\rm eq}) \nonumber \\
&& - \frac{\tau}{3} \left( \frac{w}{n_B T^2} \right) \sum_a \int d\Gamma_a \frac{p^2}{E_a}
\left( b_a - \frac{n_B}{w} E_a \right) f_a^{\rm eq} (1 - f_a^{\rm eq})
\ea
For an equilibrium Fermi-Dirac distribution function
\be
T \frac{\partial}{\partial \mu_B} f_a^{\rm eq} = b_a  f_a^{\rm eq} (1 - f_a^{\rm eq})
\ee
and
\be
T \left( T \frac{\partial}{\partial T} + \mu_B \frac{\partial}{\partial \mu_B} \right)  f_a^{\rm eq} = E_a  f_a^{\rm eq} (1 - f_a^{\rm eq})
\ee
so that the second contribution to $\lambda$ above is zero.  Hence one can use the simpler expression
\be
\lambda = \frac{\tau}{3} \left( \frac{w}{n_B T} \right)^2 \sum_a b_a \int d\Gamma_a \frac{p^2}{E_a^2}
\left( b_a - \frac{n_B}{w} E_a \right) f_a^{\rm eq} (1 - f_a^{\rm eq})
\ee
Now
\be
\sum_a b_a \int d\Gamma_a \frac{p^2}{E_a}  f_a^{\rm eq} (1 - f_a^{\rm eq}) = 3 T n_B
\ee
whereas the other integral
\bd
\sum_a b_a^2 \int d\Gamma_a \frac{p^2}{E_a^2}  f_a^{\rm eq} (1 - f_a^{\rm eq})
\ed
cannot readily be expressed in terms of thermodynamic functions.  Therefore Ref. \cite{McGill1} assumed massless particles to express the integral as
\be
\sum_a b_a^2 \int d\Gamma_a  f_a^{\rm eq} (1 - f_a^{\rm eq}) = T \chi_{\mu\mu}
\ee
which results in
\be
\kappa_B = \oneth \tau n_B \left[ \frac{T \chi_{\mu\mu}}{n_B} - \frac{3 T n_B}{w} \right]
\ee
The above formula is also true for massless particles obeying Boltzmann statistics.  For massless quarks obeying quantum statistics which all have the same chemical potential
\be
\frac{T \chi_{\mu\mu}}{n_B} = \left( \frac{1 + \mu_B^2/3 \pi^2 T^2}{1 + \mu_B^2/9 \pi^2 T^2} \right) \frac{T}{\mu_B}
\ee
whereas for massless quarks obeying Boltzmann statistics
\be
\frac{T \chi_{\mu\mu}}{n_B} = \coth\left(\frac{\mu_B}{T}\right)
\ee
It is the latter which is used in \cite{McGill1}, resulting in Eq. (\ref{kappaB}). In contrast, our expression is
\be
\kappa_B = \tau' \frac{T^4}{w} \det \chi \; v_{\sigma}^2 
= \frac{\eta}{2} \frac{T^3}{w^2} \det \chi \; v_{\sigma}^2
\ee

\bibliography{apssamp}% Produces the bibliography via BibTeX.

\end{document}